\input harvmac
 \def\quad{{\ \ }}
 \def\Bbb#1{{\fam\black\relax#1}}

\def\Bbb{\bf}

\def\ra{\r\input{unoriented.tex}}

\def\la{\langle}
\def\ra{\rangle}

\let\includefigures=\iftrue
\newfam\black
\includefigures
\input epsf
\def\figin{\epsfcheck\figin}\def\figins{\epsfcheck\figins}
\def\epsfcheck{\ifx\epsfbox\UnDeFiNeD
\message{(NO epsf.tex, FIGURES WILL BE IGNORED)}
\gdef\figin##1{\vskip2in}\gdef\figins##1{\hskip.5in}
\else\message{(FIGURES WILL BE INCLUDED)}%
\gdef\figin##1{##1}\gdef\figins##1{##1}\fi}
\def\DefWarn#1{}
\def\figinsert{\goodbreak\midinsert}
\def\ifig#1#2#3{\DefWarn#1\xdef#1{fig.~\the\figno}
\writedef{#1\leftbracket fig.\noexpand~\the\figno}%
\figinsert\figin{\centerline{#3}}\medskip\centerline{\vbox{\baselineskip12pt
\advance\hsize by -1truein\noindent\footnotefont{\bf Fig.~\the\figno:}
#2}}
\bigskip\endinsert\global\advance\figno by1}
\else
\def\ifig#1#2#3{\xdef#1{fig.~\the\figno}
\writedef{#1\leftbracket fig.\noexpand~\the\figno}%
#2}}
\global\advance\figno by1}
\fi


\def\sym{  \> {\vcenter  {\vbox
                  {\hrule height.6pt
                   \hbox {\vrule width.6pt  height5pt
                          \kern5pt
                          \vrule width.6pt  height5pt
                          \kern5pt
                          \vrule width.6pt height5pt}
                   \hrule height.6pt}
                             }
                  } \>
               }
\def\fund{  \> {\vcenter  {\vbox
                  {\hrule height.6pt
                   \hbox {\vrule width.6pt  height5pt
                          \kern5pt
                          \vrule width.6pt  height5pt }
                   \hrule height.6pt}
                             }
                       } \>
               }
\def\anti{ \>  {\vcenter  {\vbox
                  {\hrule height.6pt
                   \hbox {\vrule width.6pt  height5pt
                          \kern5pt
                          \vrule width.6pt  height5pt }
                   \hrule height.6pt
                   \hbox {\vrule width.6pt  height5pt
                          \kern5pt
                          \vrule width.6pt  height5pt }
                   \hrule height.6pt}
                             }
                  } \>
               }
\Title{\vbox{\baselineskip12pt\hbox{arXiv:0710.5170}}}
{\vbox{\centerline{Holographic Gauge Theories in Background Fields }
\vskip6pt\centerline
{And Surface Operators}}}

\centerline{Evgeny I. Buchbinder\foot{evgeny@perimeterinstitute.ca}, Jaume Gomis\foot{jgomis@perimeterinstitute.ca} and Filippo Passerini\foot{fpasserini@perimeterinstitute.ca}}
\medskip\smallskip

\bigskip\centerline{\it Perimeter Institute for Theoretical Physics}
\centerline{\it Waterloo, Ontario N2L 2Y5, Canada$^{123}$}

\vskip .03in

\bigskip\centerline{\it Department of Physics and Astronomy}
\centerline{\it University of Waterloo,  Ontario N2L 3G1, Canada$^3$}
\vskip .1in

\centerline{{\bf Abstract}}
\vskip .1in

We construct a new class of supersymmetric surface operators in ${\cal N}=4$ SYM and find the corresponding dual supergravity solutions.
 We show  that the insertion of the surface operator -- which is given by a WZW model supported on the surface -- appears  by integrating out the localized degrees of freedom along the surface which arise microscopically from  a D3/D7 brane intersection.
Consistency requires constructing ${\cal N}=4$ SYM in the D7 supergravity background and not in flat space.  This enlarges the class of holographic gauge theories dual to string theory backgrounds to gauge theories in non-trivial supergravity backgrounds. The dual Type IIB supergravity solutions we find reveal -- among other features -- that the holographic dual gauge theory does indeed live in the D7-brane background.

\Date{10/2007}


\lref\Yau{
B. R. Greene, A. D. Shapere, C. Vafa and S.-T. Yau, 
`` Stringy Cosmic Strings and Noncompact Calabi-Yau Manifolds,''
Nucl.Phys.B337:1,1990.}

\lref\Green{
G. W. Gibbons, M. B. Green and M. J. Perry, 
``Instantons and seven-branes in type IIB superstring theory,''
Phys.Lett. B370 (1996) 37-44, arXiv:hep-th/9511080.}

\lref\Berg{
E. A. Bergshoeff, J. Hartong, T. Ortin and D. Roest,
``Seven-branes and Supersymmetry,''
JHEP 0702 (2007) 003, arXiv:hep-th/0612072.}

\lref\VanProeyentwo{
L. Martucci, J. Rosseel, D. Van den Bleeken and A. Van Proeyen, 
``Dirac actions for D-branes on backgrounds with fluxes,''
Class.Quant.Grav. 22 (2005) 2745-2764, arXiv:hep-th/0504041.}
\lref\Silvaone{
D. Marolf, L. Martucci and P. J. Silva, 
``Fermions, T-duality and effective actions for D-branes in bosonic backgrounds,''
JHEP 0304 (2003) 051, arXiv:hep-th/0303209.}

\lref\Silvatwo{
D. Marolf, L. Martucci and P. J. Silva,
``Actions and Fermionic symmetries for D-branes in bosonic backgrounds,''
JHEP 0307 (2003) 019, arXiv:hep-th/0306066.}

\lref\Nicolai{
H. Nicolai, E. Sezgin and Y. Tanii,
``Conformally invariant supersymmetric field theories on $S^p \times S^1$ and super $p$-branes,''
Nucl.Phys.B305:483,1988.}

\lref\Okuyama{
K. Okuyama, 
``N=4 SYM on $R$ times $S^3$ and PP-Wave,''
JHEP 0211 (2002) 043, arXiv:hep-th/0207067.}

\lref\Pope{
H. Lu, C.N. Pope and J. Rahmfeld,
``A Construction of Killing Spinors on $S^n$,''
J.Math.Phys. 40 (1999) 4518-4526 [arXiv:hep-th/9805151].}

\lref\Claus{
P. Claus and R. Kallosh,
``Superisometries of the $AdS \times S$ Superspace,''
JHEP 9903 (1999) 014 [arXiv:hep-th/9812087].}


\newsec{Introduction and Conclusion}


The phase structure of a gauge theory can be probed by studying the  behaviour of the order parameters of the theory as we change external parameters, such as the temperature. In order to characterize  the possible phases,  one  may insert an infinitely 
heavy  probe charged particle, and study  its  response, as it will depend on the phase  the gauge theory is in. Known examples of operators inserting such probes
 are Wilson, Polyakov and 't Hooft operators, which distinguish between the confined, deconfined and  the Higgs phase.

It is  a natural question to ask whether one can construct an operator  which inserts a probe string instead of a probe particle. If so, we can then study the response of the string and analyze whether  new phases of gauge theory can be found that are not discriminated by  particle probes. Candidate probe strings range from cosmic strings to the wrapped $D$-branes of string theory. 

Geometrically, an  operator inserting a probe string is characterized by a surface $\Sigma$ in space-time, which corresponds to the worldsheet spanned by the string. One may refer to such operators as surface operators and will label them by ${\cal O}_\Sigma$.  Such operators are nonlocal in nature and the challenge is to construct them and to understand their physical meaning. For early studies of these operators see 
\lref\AlfordYX{
  M.~G.~Alford, K.~M.~Lee, J.~March-Russell and J.~Preskill,
  ``Quantum field theory of non-Abelian strings and vortices,''
  Nucl.\ Phys.\  B {\bf 384}, 251 (1992)
  [arXiv:hep-th/9112038].
}
\lref\RohmXK{
  R.~M.~Rohm,
  ``Some Current Problems In Particle Physics Beyond The Standard Model,''
  Ph.D. thesis.
}
\AlfordYX, \RohmXK.

Recently, a class of supersymmetric surface operators in ${\cal N}=4$ SYM have been constructed by Gukov and Witten  
\lref\KapustinPK{
  A.~Kapustin and E.~Witten,
  ``Electric-magnetic duality and the geometric Langlands program,''
  arXiv:hep-th/0604151.
} 
\lref\GukovJK{
  S.~Gukov and E.~Witten,
  ``Gauge theory, ramification, and the geometric langlands program,''
  arXiv:hep-th/0612073.
} 
\GukovJK\foot{These operators play an important role in enriching the  
gauge theory approach  \KapustinPK\ to the geometric Langlands program to the case with ramification.}, while the corresponding gravitational description  in terms of smooth  solutions of Type IIB supergravity which are asymptotically $AdS_5\times S^5$ has been identified in 
\lref\GomisFI{
  J.~Gomis and S.~Matsuura,
  ``Bubbling surface operators and S-duality,''
  JHEP {\bf 0706}, 025 (2007)
  [arXiv:0704.1657 [hep-th]].
}
\GomisFI.  These operators are defined by  a path integral with  a codimension two singularity  near $\Sigma$ for the ${\cal N}=4$ SYM fields.
Therefore, these operators are of disorder type as they do not admit a 
description in terms of an operator insertion which can be written in terms of the classical fields appearing in the Lagrangian.

In this paper we construct a family of surface operators in four dimensional ${\cal N}=4$ SYM that do admit a description in terms of an operator insertion made out of the ${\cal N}=4$ SYM fields. In the standard nomenclature, they are order operators. The surface operator is obtained by inserting into the ${\cal N}=4$ SYM path integral the WZW action supported on the surface $\Sigma$ 
\eqn\insertion{
 \exp\big[ i M\Gamma_{WZW}(A)\big],}
where:\foot{We note that $\Gamma_{WZW}(A)$ differs from the conventional WZW model action by the addition of a local counterterm
which is needed to guarantee that  the operator has all the appropriate symmetries.} 
\eqn\WZWintro{\eqalign{
\Gamma_{WZW}(A)&=-{1\over 8\pi}\int_{\Sigma} dx^+dx^- \hbox{Tr}\left[ \left(U^{-1}\partial_+ U\right)\left(U^{-1}\partial_- U\right)-\left(U^{-1}\partial_+ U\right)\left(V^{-1}\partial_- V\right)\right]\cr
&-{1\over 24 \pi}\int d^3x \epsilon^{ijk}\hbox{Tr}\left[\left(U^{-1}\partial_i U\right)\left(U^{-1}\partial_j U\right)\left(U^{-1}\partial_k U\right)\right].}}
The $U(N)$ group elements  $U$ and $V$ are nonlocally related to the ${\cal N}=4$ SYM gauge field $A_\mu$  along $\Sigma$ by:
 \eqn\nonlocal{
 A_{+}=U^{-1}\partial_+ U\qquad \qquad  A_{-}=V^{-1}\partial_- V.}
 $M$ is an arbitrary positive integer which labels the level of the WZW model\foot{In the string construction of this operator $N$ denotes the number of $D3$-branes while $M$ is the number of $D7$-branes.}.

We construct these operators by considering the field theory limit of a supersymmetric $D3/D7$ brane intersection along  a two dimensional surface $\Sigma$. We find that a  consistent description of the low energy dynamics of this brane intersection requires that the gauge theory on the $D3$-branes  is written down not in flat space {\it but} in the non-trivial supergravity background created by the $D7$-branes.

In this paper we construct  this supersymmetric field theory  in the $D7$-brane supergravity background and show that if we integrate out the degrees of freedom introduced by the $D7$-branes that the net effect is to insert the operator \WZWintro\ into the gauge  theory action. The same strategy of integrating out the new degrees of freedom introduced on a brane intersection was used in
\lref\GomisSB{
  J.~Gomis and F.~Passerini,
  ``Holographic Wilson loops,''
  JHEP {\bf 0608}, 074 (2006)
  [arXiv:hep-th/0604007].
}
\GomisSB\ to construct the Wilson loop operators in ${\cal N}=4$ SYM 
and to find the bulk AdS description of a Wilson loop in an arbitrary representation of the gauge group.

The physics responsible for having to consider the gauge theory on the non-trivial supergravity background 
 is that there are  chiral fermions localized on $\Sigma$ arising from the open strings stretching  between the $D3$ and $D7$ branes. It is well known that the gauge anomalies introduced by these chiral degrees of freedom are cancelled only after the appropriate Chern-Simons terms on the $D$-brane worldvolume   are included 
 \lref\GreenDD{
  M.~B.~Green, J.~A.~Harvey and G.~W.~Moore,
  ``I-brane inflow and anomalous couplings on D-branes,''
  Class.\ Quant.\ Grav.\  {\bf 14}, 47 (1997)
  [arXiv:hep-th/9605033].
} \GreenDD. The Chern-Simons terms needed to cancel the anomalies become non-trivial
due to the presence of  the RR one-form flux produced by the $D7$-branes.
We show, however,  that  it is inconsistent to  consider only the RR background produced by the $D7$-branes. One must also take into account the non-trivial background geometry and dilaton produced by the $D7$-branes as they are of the same order as the effect produced by the RR flux. This can be seen by showing that the  gauge theory in flat space in the presence of the Chern-Simon terms does not capture the supersymmetries of the brane intersection. Therefore, we are led to consider the low energy action of $N$ D3-branes in the supergravity background produced by the intersecting $D7$-branes. The gauge theory describing the low energy dynamics preserves eight supersymmetries and is  $ISO(1,1) \times SU(4)$ invariant.

Given the construction of the surface operator in term of $D$-branes we proceed to study the bulk  Type IIB supergravity description of these surface operators.
We start by showing that there is a regime in the bulk description where the $D7$-branes can be treated as probe branes in $AdS _5\times S^5$. We show that this corresponds to the regime where 
the gauge anomaly is suppressed, the Chern-Simons term can be ignored and the gauge theory lives in flat space. This corresponds to  considering the limit where $g^2M<<1$, where $g$ is the gauge theory coupling constant. In this limit the symmetries of the gauge theory are enhanced to the $SU(1,1|4)$ supergroup.

We go beyond the probe approximation and construct the exact Type IIB supergravity solutions  that are  dual to the surface operators we have constructed\foot{The supergravity solution dual to other (defect) operators in ${\cal N}=4$ have appeared in
\lref\LinNB{
  H.~Lin, O.~Lunin and J.~M.~Maldacena,
  ``Bubbling AdS space and 1/2 BPS geometries,''
  JHEP {\bf 0410}, 025 (2004)
  [arXiv:hep-th/0409174].
}
\lref\LinNH{
  H.~Lin and J.~M.~Maldacena,
  ``Fivebranes from gauge theory,''
  Phys.\ Rev.\  D {\bf 74}, 084014 (2006)
  [arXiv:hep-th/0509235].
}
\lref\YamaguchiTE{
  S.~Yamaguchi,
  ``Bubbling geometries for half BPS Wilson lines,''
  Int.\ J.\ Mod.\ Phys.\  A {\bf 22}, 1353 (2007)
  [arXiv:hep-th/0601089].
}
\lref\LuninXR{
  O.~Lunin,
  ``On gravitational description of Wilson lines,''
  JHEP {\bf 0606}, 026 (2006)
  [arXiv:hep-th/0604133].
}
\lref\GomisCU{
  J.~Gomis and C.~Romelsberger,
  ``Bubbling defect CFT's,''
  JHEP {\bf 0608}, 050 (2006)
  [arXiv:hep-th/0604155].
}

\lref\HokerXY{
  E.~D'Hoker, J.~Estes and M.~Gutperle,
  ``Exact half-BPS Type IIB interface solutions I: Local solution and
  supersymmetric Janus,''
  JHEP {\bf 0706}, 021 (2007)
  [arXiv:0705.0022 [hep-th]].
}
\lref\HokerXZ{
  E.~D'Hoker, J.~Estes and M.~Gutperle,
  ``Exact half-BPS type IIB interface solutions. II: Flux solutions and
  multi-janus,''
  JHEP {\bf 0706}, 022 (2007)
  [arXiv:0705.0024 [hep-th]].
}
\lref\HokerFQ{
  E.~D'Hoker, J.~Estes and M.~Gutperle,
  ``Gravity duals of half-BPS Wilson loops,''
  JHEP {\bf 0706}, 063 (2007)
  [arXiv:0705.1004 [hep-th]].
}
\LinNB,\LinNH,\YamaguchiTE,\LuninXR,\GomisCU,\GomisFI,\HokerXY,\HokerXZ,\HokerFQ.}. These solutions can be found by taking the near horizon limit of the supergravity solution describing the localized $D3/D7$ brane intersection from which the surface operator is constructed. 
 The dual supergravity solutions take the form of a warped 
$AdS_3 \times S^5\times {\cal M}$ metric, where ${\cal M}$ is a two dimensional complex manifold. These solutions   also shed light on the geometry where the holographic field theory lives. One can  infer that the gauge theory lives on the curved background produced by the $D7$-branes by analyzing the dual supergravity geometry near the conformal boundary, thus showing that holography requires putting the gauge theory in a curved space-time. The explicit construction of the supergravity solutions also gives us information about the quantum properties of our surface operators. To leading order in the $g^2M$ expansion, the surface operator preserves an $SO(2,2)\subset SU(1,1|4)$ symmetry, which is associated with conformal transformations on 
the surface $\Sigma={\Bbb R}^{1,1}$. In the probe brane 
description  -- where $g^2M$ effects are suppressed -- we also have 
the $SO(2,2)$ symmetry, while the explicit supergravity solution shows that the 
 $SO(2,2)$ symmetry is broken by $g^2M$ corrections. 
 This shows  that $g^2M$ corrections in the field theory break conformal invariance, which can be seen explicitly by analyzing the gauge theory on the $D7$-brane background. This field theory statement is reminiscent\foot{Such models have been realized  in string theory using brane intersections in e.g.
 \lref\KarchSH{
  A.~Karch and E.~Katz,
  ``Adding flavor to AdS/CFT,''
  JHEP {\bf 0206}, 043 (2002)
  [arXiv:hep-th/0205236].
}
 \lref\KruczenskiBE{
  M.~Kruczenski, D.~Mateos, R.~C.~Myers and D.~J.~Winters,
   ``Meson spectroscopy in AdS/CFT with flavour,''
  JHEP {\bf 0307}, 049 (2003)
  [arXiv:hep-th/0304032].
}
\KarchSH, \KruczenskiBE. For attempts at computing the supergravity description of this system
 see e.g
 \lref\AharonyXZ{
  O.~Aharony, A.~Fayyazuddin and J.~M.~Maldacena,
  ``The large N limit of N = 2,1 field theories from three-branes in
  F-theory,''
  JHEP {\bf 9807}, 013 (1998)
  [arXiv:hep-th/9806159].
}
 \lref\GranaXN{
  M.~Grana and J.~Polchinski,
  ``Gauge / gravity duals with holomorphic dilaton,''
  Phys.\ Rev.\  D {\bf 65}, 126005 (2002)
  [arXiv:hep-th/0106014].
}
\lref\BurringtonID{
  B.~A.~Burrington, J.~T.~Liu, L.~A.~Pando Zayas and D.~Vaman,
  ``Holographic duals of flavored N = 1 super Yang-Mills: Beyond the probe
  approximation,''
  JHEP {\bf 0502}, 022 (2005)
  [arXiv:hep-th/0406207].
}
\AharonyXZ, \GranaXN, \BurringtonID.}
  to the breaking of conformal invariance by $g^2M$ effects that occurs when considering ${\cal N}=4$ SYM coupled to $M$ hypermultiplets, whose $\beta$-function is proportional to $g^2M$.
 
A general lesson that emerges from this work is that gauge theories in non-trivial supergravity backgrounds can also serve as the holographic description of string theory backgrounds. It would be interesting to explore in more detail the dictionary relating bulk and gauge theory computations.
An important problem for the future is to understand the physics encoded in the expectation value of surface operators and to determine whether they can be useful probes of new phases of gauge theory. For the surface operators in this paper it would be interesting to compute their expectation value  in perturbation theory. Given that these operators are supersymmetric it is conceivable that the computation of their expectation value can be performed in a reduced model, just like the expectation value of supersymetric circular Wilson loops can be computed by a matrix integral
\lref\EricksonAF{
  J.~K.~Erickson, G.~W.~Semenoff and K.~Zarembo,
  ``Wilson loops in N = 4 supersymmetric Yang-Mills theory,''
  Nucl.\ Phys.\  B {\bf 582}, 155 (2000)
  [arXiv:hep-th/0003055].
}
\lref\DrukkerRR{
  N.~Drukker and D.~J.~Gross,
  ``An exact prediction of N = 4 SUSYM theory for string theory,''
  J.\ Math.\ Phys.\  {\bf 42}, 2896 (2001)
  [arXiv:hep-th/0010274].
}
\EricksonAF, \DrukkerRR. One may be able to derive  the reduced model by topologically twisting the gauge theory by the supercharges preserved by the surface operator. It would also be interesting to compute the expectation value of the surface operator by calculating the on-shell action of the corresponding   supergravity solutions.

The plan  of the paper is as follows. In section $2$ we introduce the $D3/D7$ brane  intersection, the corresponding  low energy spectrum and    discuss the cancellation of the gauge anomalies via anomaly inflow. We show that the gauge theory on the $D3$-branes has to be placed in the supergravity background produced by the $D7$-branes and  construct explicitly the relevant gauge theory action, derive the appropriate supersymmetry transformations and show that the action has all the required symmetries. We integrate out all the degrees of freedom introduced by the $D7$-branes and show that the net effect is to insert the WZW action \insertion\ into the ${\cal N}=4$ SYM path integral. In section 3 we give the bulk description of the surface operators. We show that there is a regime where the $D7$-branes can be treated as probe branes in 
$AdS_5\times S^5$ and identify this with the regime in the field theory 
where the anomaly is suppressed, the Chern-Simons term can be 
ignored and the gauge theory lives in flat space. We 
find the explicit exact supergravity solution describing the supergravity background 
produced by the localized $D3/D7$ brane intersection and show that in the near horizon limit it 
is described by an $AdS_3\times S^5$ warped metric  over a two dimensional manifold. 
We show that the metric on the boundary, where the gauge theory lives, is precisely the 
$D7$-brane metric on which we constructed the field theory in section 2. Some of technical 
details  and computations are relegated to the Appendices.

\noindent
{\it Note Added}: While this work was being completed, the paper 
\lref\HarveyAB{
  J.~A.~Harvey and A.~B.~Royston,
  ``Localized Modes at a D-brane--O-plane Intersection and Heterotic Alice
  Strings,''
  arXiv:0709.1482 [hep-th].
} \HarveyAB\ appeared, which also discusses the gauge theory action on the $D3/D7$  brane intersection and overlaps with   section $2$.


\newsec{Gauge Theory and Surface Operators}


\subsec{Brane Intersection and  Anomalies}


The surface operators   in this paper are constructed from the low energy field theory on a $D3/D7$ 
brane configuration that intersects along a surface 
$\Sigma={\Bbb R}^{1,1}$. More precisely,  we consider the effective description on $N$ 
$D3$-branes with worldvolume 
coordinates $x^\mu=(x^0,x^1,x^2,x^3)$ and $M$ 
$D7$-branes whose worldvolume is parameterized by $(x^0,x^1)$ 
and $x^I=(x^4,x^5,x^6,x^7,x^8,x^9)$. The coordinates that parametrize  
the surface $\Sigma$ are $x^0$ and $x^1$: 
\smallskip
\eqn\braneconf{\matrix{\ \ &0&1&2&3&4&5&6&7&8&9\cr
N\,D3&\hbox{X}&\hbox{X}&\hbox{X}&\hbox{X}&&&&&\cr
M\,D7&\hbox{X}& \hbox{X}&&& \hbox{X}&\hbox{X}&\hbox{X}&\hbox{X}&\hbox{X}&\hbox{X}}}
\medskip
\noindent
The  supersymmetries preserved by the 
 $D3$-branes are  the following\foot{In this paper we denote the $\gamma$-matrices in flat space by 
$\gamma$. The curved 
 space $\gamma$-matrices are denoted by $\Gamma$. They satisfy $\{\Gamma^M,\Gamma^N\}=2g^{MN}$, where $g_{MN}$ is the space-time metric.} 
 \eqn\dths{i\gamma^{0123}\epsilon=\epsilon,}
where $\epsilon$ is a ten dimensional complex Weyl spinor  satisfying
$\gamma^{01\ldots89}\epsilon=\epsilon$, which labels the thirty-two supersymmetries of Type IIB supergravity. The supersymmetries preserved by the $D7$-branes are given by:
\eqn\dses{i\gamma^{01456789}\epsilon=\epsilon.}
Therefore, in total there are eight supersymmetries   preserved by the brane intersection, which can be shown to be chiral in the two dimensional intersection. 
If we introduce  coordinates
 \eqn\coor{x^\pm=x^0\pm x^1\qquad z=x^2+ix^3,} 
 then the unbroken  supersymmetries   satisfy
 \eqn\susy{\gamma_+\epsilon=0,}
or can alternatively be written as 
\eqn\susyz{\gamma_{\bar z}\epsilon=0,}
where:
\eqn\gp{\gamma_+={1\over 2}(\gamma_0+\gamma_1), \qquad \gamma_{\bar z}={1\over 2}(\gamma_2+i\gamma_3).}

In constructing  the supersymmetry transformations of the gauge theory living on the brane intersection we will use four 
dimensional Weyl spinors. In the four dimensional notation, the sixteen 
supersymmetries preserved by the  $D3$-branes \dths\  are generated by 
$(\epsilon_\alpha{}^i, \bar{\epsilon}_{\dot{\alpha}i})$, 
where   $\epsilon_\alpha{}^i$ is a four dimensional Weyl 
spinor of positive chirality transforming in the $({\bf 2},{\bf 4})$ 
representation of 
$SL(2,{\Bbb C})\times SU(4)$ and $\bar{\epsilon}_{\dot{\alpha}i}=(\epsilon_\alpha{}^i)^*$. 
These spinors generate the usual   Poincare supersymmetry transformations of ${\cal N}=4$ Yang-Mills theory. 
In this notation, the projectors \susy\ and~\susyz\ can be written as:\foot{Our conventions on $\sigma$-matrices are summarized in Appendix A. They are essentially the same as those  in the book 
\lref\BuchbinderQV{
  I.~L.~Buchbinder and S.~M.~Kuzenko,
  ``Ideas and methods of supersymmetry and supergravity: Or a walk through
  superspace,''
{\it  Bristol, UK: IOP (1998) 656 p}
} \BuchbinderQV.} 
\eqn\four{{\tilde\sigma}_{+}{}^{{\dot \alpha}\alpha }\epsilon_\alpha{}^i=0,\qquad 
{\tilde\sigma}_{{\bar z}}{}^{{\dot \alpha}\alpha }\epsilon_\alpha{}^i=0.}
Therefore, the projections \four\ imply that $\epsilon_1{}^i=\epsilon_2{}^i$, which parametrize the eight real supersymmetries preserved by the brane intersection.

In the low energy limit -- where $\alpha'\rightarrow 0$ --  massive open strings and  closed string excitations decouple and  only the  massless open strings are relevant. The 3-3 strings yield the spectrum  of four dimensional ${\cal N}=4$ SYM while the quantization of the 3-7 open strings results in two dimensional chiral fermions $\chi$ localized on the intersection, and  transform in the $(N,{\bar M})$ representation of $U(N)\times U(M)$.  The massless 7-7 strings give rise to a SYM multiplet in eight dimensions, but these degrees of freedom are non-dynamical in the decoupling limit and appear in the effective action only as Lagrange multipliers.  
 
  The action for the localized chiral fermions is given by 
 \eqn\defe{S_{defect}=\int dx^+ d x^-\ {\bar \chi}(\partial_++A_++{\tilde A}_+)\chi,}
 where $A$ and $\tilde{A}$ denote the $D3$ and $D7$-brane gauge fields respectively 
and we have used the coordinates introduced in \coor. Of the usual Poincare supersymmetries of ${\cal N}=4$ SYM, whose relevant transformations are given by
 \eqn\susyvar{\delta A_\mu=-i{\bar \lambda}_{{\dot \alpha}i} 
{\tilde\sigma}_{\mu}{}^{{\dot \alpha}\alpha }\epsilon_\alpha{}^i +{\rm c.c.}, 
\qquad \delta\chi=0,\qquad \delta {\tilde A}_\mu=0,}
the defect term \defe\ is invariant under  those supersymmetries for which
$\delta A_+=0$, which are precisely the ones that satisfy the projections in \four\ arising from the 
$D3/D7$ brane intersection.

Quantum mechanically, the path integral over the localized chiral fermions $\chi$ is not well defined due to the presence of gauge anomalies in the intersection. In order to see how to cure this problem, 
it is convenient to split the $U(N)$ and $U(M)$ gauge fields  into   $SU(N)\times U(1)$ and $SU(M)\times  U(1)$ gauge fields .  With some abuse of notation, we   denote the  $SU(N)$ and $SU(M)$  parts of the gauge field  by $A$ and ${\tilde A}$ respectively, while the corresponding $U(1)$ parts of the gauge field are denoted by $a$ and ${\tilde a}$.
Then,
the variation of the quantum effective action under an  $SU(N)\times SU(M)$ gauge transformation  
\eqn\gauvar{\delta A_\mu=\partial_\mu L+[A_\mu, L],\qquad  \delta {\tilde A}_\mu=\partial_\mu {\tilde L}+[{\tilde A}_\mu, {\tilde L}]}
is given by 
\eqn\aly{\delta_{L,{\tilde L}}S={1\over 8 \pi}\int d x^{+} d x^{-}\left[M\hbox{Tr}_{SU(N)}(LdA)+N\hbox{Tr}_{SU(M)}({\tilde L}d{\tilde A})\right],}
so that the theory is anomalous under $SU(N)\times SU(M)$ gauge  transformations.
Likewise, $U(1) \times U(1)$  gauge  transformations
\eqn\onevar{\delta A_\mu = \partial_\mu l,\qquad \delta {\tilde A}_\mu = \partial_\mu {\tilde l},}
on the quantum effective action yield
\eqn\alyone{\delta_{l,{\tilde l}}S={1\over 8 \pi}\int d x^{+} d x^{-}NM(l-{\tilde l})(f_{+-}-{\tilde f}_{+-}),}
 so that the theory is anomalous under the $U(1)$ gauge transformations generated by $l-\tilde{l}$, 
and where: 
\eqn\onef{f=da,\qquad {\tilde f}=d{\tilde a}.}

Anomalies supported on  $D$-brane intersections are cancelled by  the anomaly 
inflow mechanism  \GreenDD, which relies on the presence of Chern-Simons 
couplings in the $D$-brane worldvolume. The Chern-Simons terms that couple to 
the  $SU(N)$ and $SU(M)$ gauge fields are given by 
\eqn\cs{S_{CS}(A)=-{(2 \pi \alpha^{\prime})^2 \tau_3 \over 2}
\int G_1\wedge \hbox{Tr}\left(A\wedge d A +{2\over 3}A\wedge A \wedge A\right)}
and 
\eqn\cst{S_{CS}({\tilde A})=-{(2 \pi \alpha^{\prime})^2 \tau_7 \over 2}
\int G_5\wedge \hbox{Tr}
\left({\tilde A} \wedge d {\tilde A}+{2\over 3}{\tilde A}\wedge{\tilde A} \wedge {\tilde A}\right),}
where $g_s$ is the string coupling constant and  $\tau_3$ and $\tau_7$ is the $D3$  and $D7$-brane
tension respectively:
\eqn\newone{
\tau_3={1 \over g_s (2 \pi)^3 \alpha^{\prime 2}}, 
\qquad
\tau_7={1 \over g_s (2 \pi)^7 \alpha^{\prime 4}}.}
$G_1$ 
is the RR one-form flux produced by the stack of  $D7$-branes 
and $G_5$ is the self-dual RR five-form flux produced by the stack of $D3$-branes. 

In the presence of localized $D$-brane sources, 
the Bianchi identities for the RR fields are modified in a way 
that the Chern-Simons terms become non-trivial. 
In our case, the modified Bianchi identities are given by
\eqn\bi{
dG_1= M G_{10} \tau_7 \delta^2 (z \bar z)=
g_sM  \delta^2(z \bar z)}
and 
\eqn\bit{dG_5= N G_{10} \tau_3  \delta(x^4)\delta(x^5)\ldots\delta(x^9),}
where $G_{10}$ is the ten-dimensional Newton's constant which is given by:
\eqn\newtwo{
G_{10}=g_s^2 (2 \pi)^7 \alpha^{\prime 4}.}
Therefore,  under an $SU(N)\times SU(M)$ gauge 
transformation \gauvar, the Chern-Simons terms \cs and \cst\  
are not invariant, and reproduce the two-dimensional anomaly
\eqn\varcs{\delta S_{CS}(A)+\delta S_{CS}({\tilde A})=-{1\over 8 \pi}
\int d x^{+} d x^{-}\left[M\hbox{Tr}_{SU(N)}(LdA)+N\hbox{Tr}_{SU(M)}
({\tilde L}d{\tilde A})\right],}
where  $L$ and ${\tilde L}$ are taken to vanish at infinity. 
This mechanism provides a cancellation of the $SU(N)$  and $SU(M)$ gauge anomalies \GreenDD.  

The Chern-Simons terms containing the $U(1)$ gauge fields $a$ and ${\tilde a}$ are more involved. 
They have been studied in 
\lref\ItzhakiTU{
  N.~Itzhaki, D.~Kutasov and N.~Seiberg,
  ``I-brane dynamics,''
  JHEP {\bf 0601}, 119 (2006)
  [arXiv:hep-th/0508025].
} \ItzhakiTU,
where the anomalies of a closely related $D5$/$D5$ brane intersection along a 
two dimensional defect were 
studied\foot{The physics of that system is quite different from the $D3$/$D7$ 
system studied in this paper. In \ItzhakiTU\ it was argued that the dynamics of the gauge fields pushes the fermions away from the intersection by a distance determined by the (dimensionful) gauge theory coupling constant.
 In our system the fermions are stuck at the intersection since the 
$U(N)$ coupling constant is dimensionless unlike the one on the $D5$-branes which is 
dimensionful while the $U(M)$ gauge coupling constant vanishes in the decoupling limit, 
pinning down the fermions at the intersection. Moreover, in  
\ItzhakiTU\ the symmetry is enhanced from $ISO(1,1)$ to $ISO(1,2)$ while in our 
system the symmetry is enhanced from $ISO(1,1)$ to $SO(2,2)$, but only to 
leading order in the $g^2M$ expansion. Here we also resolve a 
puzzle left over in their paper, which is to construct 
the gauge theory action with all the expected supersymmetries.}.
The analogous terms for the  
$D3$/$D7$ system are given by: 
\eqn\csone{\eqalign{S_{CS}(a, {\tilde a})=&-{(2 \pi \alpha^{\prime})^2 \tau_3\over 2} 
N \int G_1\wedge a \wedge f-
{(2 \pi \alpha^{\prime})^2 \tau_7
\over 2} M \int G_5\wedge {\tilde a} \wedge {\tilde f}\cr +&{(2 \pi \alpha^{\prime})^2 \tau_3 \over 2} N 
\int G_1\wedge a \wedge {\tilde f}+{(2 \pi \alpha^{\prime})^2 \tau_7\over 2} M 
\int G_5\wedge {\tilde a} \wedge f.}}
The first two terms are the usual Chern-Simons couplings  analogous to \cs\ and \cst.  
The third term arises from the familiar  coupling on the $D3$-brane worldvolume of the form 
\eqn\csterm{\int a\wedge F_3,}
 where $F_3$ is the RR three-form flux, which as argued in \ItzhakiTU\ is   given by  $F_3=G_1\wedge \tilde{f}$ in the presence of $G_1$ and  $\tilde{f}$ background fields.   
Note that $\tilde{f}$ in the third term is to be evaluated at $x^I=0$. 
 Similarly, the last term arises from  the Chern-Simons coupling on the $D7$-brane 
 \eqn\cstermb{\int {\tilde a}\wedge F_7,} 
 where the RR seven-form flux is now  given 
by $G_5\wedge f$, where   $f$ is to be evaluated at $z=0$.
If we now perform a $U(1)\times U(1)$ gauge transformation, the  
variation of  \csone\ is given by
\eqn\vcsone{\delta S_{CS}(a, {\bar a})=-{1\over 8 \pi}
\int d x^{+} d x^{-}NM(l-{\tilde l})(f_{+-}-{\tilde f}_{+-}),}
where we have used the modified Bianchi identities  \bi\ and \bit. Therefore, by including all the Chern-Simons couplings all anomalies cancel.
\medskip
\medskip
\noindent
{\it  Field Theory Construction of Gauge Theories with Anomaly Inflow}
\medskip
\medskip

Turning on the RR fluxes \bi\ and \bit\ produced by the $D3$ and $D7$
 branes is 
crucial in obtaining an effective theory  which is anomaly free. Usually, in analyzing the low energy gauge theory on a $D$-brane intersection in flat space we can ignore the RR flux produced by the branes. However, whenever there are localized gauge  anomalies the RR flux cannot be neglected as it generates the required  Chern-Simons needed to cancel the anomaly.  But $D$-branes also source other supergravity fields, such as the metric and the dilaton. It is therefore inconsistent to study the low energy gauge theory in flat space with only the addition of the RR-induced Chern-Simons terms. Physically, one must consider the gauge theory in the {\it full} supergravity background produced by the other $D$-brane, as the effect of the metric and dilaton is of the same order as the effect of the RR flux. 

One way to see that it is inconsistent to consider the gauge theory on the $D3$-branes in flat space and in the presence of only the RR-flux produced by the $D7$-branes is to note that  the naive action of the system
\eqn\naive{
S=S_{{\cal N}=4}+S_{defect}+S_{CS}(A)+S_{CS}({\tilde A})+S_{CS}(a,\tilde{a}),}
is not supersymmetric, where $S_{{\cal N}=4}$ is the usual flat space action of ${\cal N}=4$ SYM and the other terms appear in \defe, \cs, \cst\ and \csone\ respectively. 
In particular, this low-energy gauge theory does not capture the 
supersymmetries of the brane intersection \four, and therefore is not a 
faithful description of the low energy dynamics. 

In the rest of this section we construct the low energy gauge theory living on the $D3$-branes when embedded in the full supergravity background of the $D7$-branes -- which includes the appropriate Chern-Simons terms -- and show that the field theory has all the required symmetries.




\subsec{The $D7$-Brane Background}


As   just argued, we must construct the low energy gauge theory on the $D3$-branes when placed in the full supergravity background of the $D7$-branes. We will devote this subsection to reviewing the salient features of the $D7$-brane background.

 The  metric
produced by the $D7$-branes in the brane array \braneconf\ is given by
\eqn\zerotwoone{ds^2=g_{MN}dx^Mdx^N=H_7^{-1/2}(-(dx^0)^2+(dx^1)^2+ dx^I dx^I)+H_7^{1/2}dz d{\bar z},}
where the coordinates are defined in \braneconf.
The RR axion $C$ and the dilaton $\Phi$ can be combined into a complex field $\tau$ 
with is holomorphic in $z$, so that the axion and the  
dilaton produced by the $D7$-branes is given by:
\eqn\zerotwotwo{\eqalign{
\partial_{{\bar z}}\tau&=0\qquad \hbox{where}\qquad \tau=C+ie^{-\Phi}
\cr
e^{-\Phi}&=H_7.}}

This background solves the Killing spinor equations of Type IIB supergravity 
\eqn\zerotwothree{\eqalign{
&\delta\Psi_M=\partial_M \epsilon +{1\over 4}\omega_{M}^{AB}\Gamma_{AB} \epsilon
-{i\over 8}e^{\Phi}\partial_N C \Gamma^N\Gamma_M\epsilon =0,\cr 
&\delta \psi =(\Gamma^M\partial_M\Phi)\epsilon 
+i e^{\Phi}\partial_M C \Gamma^M \epsilon=0,}}
and preserves the sixteen supersymmetries  satisfying
\eqn\constraintspinor{
\epsilon=H_7^{-1/8}\epsilon_0, \quad \gamma_{\bar z}\epsilon_0=0,}
where $\Psi_M$ and $\psi$ are the ten-dimensional gravitino and dilatino 
respectively. 

The simplest solution describes the local fields around a  coincident stack of $D7$-branes. This local  solution has a $U(1)$ symmetry,
which acts by rotations in the space transverse to the $D7$-branes, 
which is parametrized by the coordinate $z$. It is given by
\eqn\zerotwofour{\tau=i\tau_0+{g_s M\over 2 \pi i}\ln z,}
so that 
\eqn\fieldsaround{
 \quad
e^{-\Phi}=H_7=\tau_0 -{g_s M\over 2\pi}\ln r, \quad C={g_s M\over 2\pi}\theta,}
where $z=re^{i\theta}$ and $\tau_0$ is an arbitrary real constant. 
This solution, however, is only valid very near the branes -- for small $r$ -- as  $e^{-\Phi}$ becomes negative at a finite distance and we encounter a singularity. The local solution for separated branes corresponds to 
\eqn\separa{
\tau=i\tau_0+{g_s\over 2 \pi i}\sum_{l=1}^M\ln (z-z_l),}
where $z_l$ is location of the $l$-th $D7$-brane.

As shown in \Yau\ (see \Green, \Berg\ for more recent discussions), the local 
solution can be patched into   global solutions that avoid the pathologies of the local one. 
The global solutions break the $U(1)$ symmetry present in the local solution of coincident $D7$-branes.
In order to describe them it is convenient to switch to the Einstein frame, where  
the $SL(2, {\Bbb Z})$ invariance of Type IIB string theory is manifest. 
In this frame, the local metric is given by
\eqn\zerotwofive{ds^2=-(dx^0)^2+(dx^1)^2+ dx^I dx^I+H_7 dz d{\bar z}.}
Since $\tau$ is defined up to the action of  $SL(2, {\Bbb Z})$  and $Im\ \tau >0$,
it follows that $\tau$ takes values in the fundamental domain 
${\cal F}={\cal H}^+ /SL(2, {\Bbb Z})$, where ${\cal H}^+$ is the 
upper half plane. In order to find a global solution for $\tau$ one has to consider 
the one-to-one map $j:{\cal F}\to {\Bbb C}$ from the fundamental domain ${\cal F}$   to 
the   complex plane $\Bbb{C}$. This map $j$ is well-known and given by
\eqn\zerotwosix{j(\tau)={(\theta_2(\tau)^8 + \theta_3(\tau)^8+\theta_4(\tau)^8)^3
\over\eta(\tau)^{24}},}
where the $\theta$'s are the usual theta-functions while  $\eta$ is the Dedekind $\eta$-function 
\eqn\zerotwoseven{\eta(\tau)=q^{1/24}\prod_{n} (1-q^n),}
where $q=e^{2\pi i \tau}$. 
Then the various solutions for $\tau$ are given by 
\eqn\zerotwoeight{j(\tau(z))=g(z),}
where $g(z)$ is an arbitrary meromorphic function in the complex plane. 
For a stack of $M$ coincident $D7$-branes we have 
\eqn\zerotwoeightpointone{g(z)=a+{b\over z^{g_s M}},}
where $a$ sets the value of the dilaton at infinity and $b$ is related to 
$\tau_0$ in \fieldsaround. Indeed, for  $Im\ \tau>>1$, $j(\tau)\simeq e^{-2 \pi i \tau}$ which implies the 
local behavior \zerotwofour\ near $z=0$. 

In general, different choices of $g(z)$
correspond to different types of $D7$-brane solutions. The metric can be written in the 
following form
\eqn\zerotwonine{ds^2=-(dx^0)^2+(dx^1)^2+ dx^I dx^I+H_7 f\bar fdz d{\bar z},}
where  as in the local case 
$H_7=e^{-\Phi}$  and where $f$ is a holomorphic function of $z$. 
Locally, one can always choose a coordinate system where 
$f\bar f dz \bar z = dz^{\prime} d{\bar z}^{\prime}$ for some
local coordinates $z^{\prime}$ and ${\bar z}^{\prime}$. This brings 
the metric~\zerotwonine\ to the usual local form~\zerotwofive.
However, globally this cannot be done as discussed above. 
For the metric to be globally defined, $H_7 f \bar f$ has to be 
$SL(2, {\Bbb Z})$ invariant. The solution studied in~\Yau\ is given by
\eqn\zerotwoten{
H_7 f \bar f=e^{-\Phi}\eta^2 {\bar \eta}^2 
|\prod_{i=1}^M (z-z_i)^{-1/12}|^2,}
where $z_i$'s are the location of the  poles of $g(z)$,  which correspond to the position   of the various $D7$-branes in the $z$-plane\foot{There are restrictions on the range of $M$ coming from the fact that 
for $M$ large enough the space becomes compact. This was studied in
detail in~\Yau. We will not discuss this point in this paper.}.

The metric~\zerotwonine\ is smooth everywhere except 
1) at $z=z_i$ where it behaves as $\ln |z-z_i|$ due to the presence of a $D7$-brane source there and 
2) at infinity, where it has a conical singularity with deficit angle 
$\delta ={\pi M\over 6}$. In this paper, we will mostly be using 
the $D7$-brane background in the local form~\zerotwoone, \zerotwotwo.
However, as we explained the generalization to the global case is straightforward. 

We finish this subsection by constructing the Killing spinors of the gauge theory on the $D3$-branes when placed in the background of the $D7$-branes. If we consider the $D3/D7$ intersection in \braneconf, we need the restriction of the 
$D7$-brane background to the worldvolume of the $D3$-branes. Then the induced metric on the $D3$-branes is given by:
\eqn\zerotwoeleven{ds^2=g_{\mu \nu}dx^{\mu} dx^{\nu}=
-H_7^{-1/2}dx^+ dx^- + H_7^{1/2} dz d {\bar z}.}
The Killing spinor equation satisfied by the four dimensional spinors 
$\epsilon_{\alpha}{}^i$ that generate the worldvolume supersymmetry 
transformations on the $D3$-branes  is given by\foot{To go to the four dimensional notation we have used: 
\eqn\gammamatrix{\Gamma^{\mu}=i\left(\eqalign{&0\qquad \sigma^\mu \cr  & 
{\tilde \sigma}^\mu \quad\quad 0 }\right).}} 
\eqn\zerotwotwelve{\eqalign{
&D_{\mu}\epsilon_{\alpha}{}^i =-{i\over 8}
e^{\Phi} \partial_{\nu}C \sigma^{\nu}_{{\ }\alpha \dot\alpha}
\tilde{\sigma}_{\mu}^{{\ }\dot\alpha \beta} \epsilon_{\beta}{}^i},}
where $D_{\mu}$ is the covariant derivative in the background metric \zerotwoeleven. Therefore
\eqn\formkilling{
\epsilon_{\alpha}{}^i=H_7^{-1/8}\epsilon_{0{} \alpha}{}^i,}
where 
\eqn\constrapinor{
 \tilde{\sigma}^{{\ }\dot\alpha \alpha}_{\bar z}\epsilon_{\alpha}{}^i=0, \quad\qquad
\tilde{\sigma}^{{\ }\dot\alpha \alpha}_{+}\epsilon_{\alpha}{}^i=0,}
thus reproducing the supersymmetry conditions derived for the brane intersection \four. In the next subsection we write down the action   and supersymmetry transformations of the $D3/D7$ low  energy gauge theory and show that the preserved Killing spinors  satisfy 
\zerotwotwelve\ subject to the constraints \constrapinor.

%

\subsec{Holographic Gauge Theory in Background Fields}


In this subsection we construct the low energy gauge theory on the $D3$-branes 
when placed in the full supergravity background of the $D7$-branes. This is the appropriate  decoupled field theory that 
holographically describes the physics of the dual closed string background, 
which we obtain in section 3 by finding the supergravity solution of 
the $D3/D7$ intersection. We also construct the corresponding supersymmetry 
transformations and show that the action is  invariant under the subset 
of ${\cal N}=4$ supersymmetry transformations satisfying the 
restrictions \zerotwotwelve\ and \constrapinor, which are precisely the supersymmetries preserved by the $D$-brane intersection in flat space \braneconf.

There is a systematic way of constructing the action and supersymmetry transformations  
on a single $D$-brane in an arbitrary 
supergravity background. The starting point is to consider the covariant  
$D$-brane action in an arbitrary curved superspace background
\lref\CederwallRI{
  M.~Cederwall, A.~von Gussich, B.~E.~W.~Nilsson, P.~Sundell and A.~Westerberg,
  ``The Dirichlet super-p-branes in ten-dimensional type IIA and IIB
  Nucl.\ Phys.\  B {\bf 490}, 179 (1997)
  [arXiv:hep-th/9611159].
}
\lref\BergshoeffTU{
  E.~Bergshoeff and P.~K.~Townsend,
  ``Super D-branes,''
  Nucl.\ Phys.\  B {\bf 490}, 145 (1997)
  [arXiv:hep-th/9611173].
}
\CederwallRI, \BergshoeffTU\ (which generalizes the flat space construction in 
\lref\AganagicNN{
  M.~Aganagic, C.~Popescu and J.~H.~Schwarz,
  ``Gauge-invariant and gauge-fixed D-brane actions,''
  Nucl.\ Phys.\  B {\bf 495}, 99 (1997)
  [arXiv:hep-th/9612080].
}
\AganagicNN.)
These actions can in principle be expanded to all orders in the fermions around a given background, even though explicit formulas are not easy to obtain. The covariant action  has $\kappa$-symmetry and diffeomorphism invariance. By fixing $\kappa$-symmetry we can gauge away sixteen of the thirty two fermions of Type IIB supergravity superspace. The remaining sixteen fermions are then identified with the gauginos filling up the  SYM multiplet living on the $D$-brane. Likewise, worldvolume diffeomorphisms can be fixed by specifying how the brane is embedded in the background, which allows for the identification of the scalars of the SYM multiplet parametrizing the position of the $D$-brane. 

In order to construct the explicit supersymmetry transformations of the gauge fixed action one must combine the superspace supersymmetry transformations on the physical fields together with a compensating  $\kappa$ and diffeomorphism transformation to preserve the gauge fixing condition.

Since we are interested in considering a decoupling limit, where $\alpha^\prime\rightarrow 0$, this procedure simplifies considerably. In this limit the only terms in the action that survive are quadratic in the fields. Fortunately, the explicit expression for  the $D$-brane action to quadratic order in the fermions in an arbitrary supergravity background  
can be found in \VanProeyentwo\foot{In that paper the action is written to 
quadratic order in the fermionic fields and to all order in the bosonic fields. 
In the decoupling limit, we will only need to extract the action to 
quadratic order in the bosonic fields.} (see also  \Silvaone,\Silvatwo,
\lref\BandosWB{
  I.~Bandos and D.~Sorokin,
  ``Aspects of D-brane dynamics in supergravity backgrounds with fluxes,
  Nucl.\ Phys.\  B {\bf 759}, 399 (2006)
  [arXiv:hep-th/0607163].
}\BandosWB ).
 This approach gives
the brane action
quadratic in fermions with fixed $\kappa$-supersymmetry and diffeomorphisms
in an arbitrary supersymmetric background. 
Therefore, we  start by finding the action for a 
single $D3$-brane in the $D7$-brane background following \VanProeyentwo. 
Later we will show how to extend this analysis to the case when the gauge group is non-Abelian.

Let us start with the bosonic action in the $D7$-brane background. The action for the gauge field
$A_{\mu}$ is straightforward to write down. It is given by
\eqn\zerotwothirteen{
S_V=-{T_3 \over 4}\int d^4 x \sqrt{-g} e^{-\Phi}F_{\mu \nu}F^{\mu \nu}
-{T_3 \over 4}\int d^4 x \sqrt{-g}\partial_{\mu}C 
\epsilon^{\mu \nu \rho \sigma}A_{\nu} F_{\rho \sigma},}
where 
\eqn\zerotwothirteenpointone{T_3=(2 \pi \alpha^{\prime})^2 \tau_3=
{1 \over 2 \pi g_s}={1\over g^2},}
where $g$ is the SYM coupling constant.
The coordinates $x^\mu=(x^+,x^-,z,{\bar z})$ describe the coordinates 
along the $D3$-brane worldvolume as defined in \braneconf. 
The metric used on the $D3$-brane worldvolume is the induced 
metric \zerotwoeleven\ from the $D7$-brane background. 

In order to obtain the action for the scalar fields on the $D3$-brane 
it is important to properly identify which are the fields describing the $D$-brane fluctuations. We introduce vielbeins which are adapted to the symmetries preserved by the $D3$-brane $(e^{\hat{\mu}}, e^{\hat{I}})$, where $\hat{\mu}$ and $\hat{I}$   denote the flat indices
along and transverse to the $D3$-brane respectively. The static gauge is fixed by the requirement that the pullback
of the vielbein $e^{\hat{I}}_{I}$  on the $D3$-brane vanishes and the pullback of the 
vielbein $e^{\hat{\mu}}_{\mu}$ forms a $D3$-brane worldvolume 
vielbein. The physical scalar fields are parametrized by 
\eqn\zerotwoforteenpointone{
\varphi^{\hat{I}}=e^{\hat{I}}_{I} \delta x^I}
rather than by the fluctuations in the transverse coordinates
$\delta x^I$. The scalar fields $\varphi^{\hat{I}}$ transform  under the local tangent space $SU(4)\simeq SO(6)$ symmetries while the fluctuations $\delta x^I$ transform under diffeomorphisms in the transverse space.
This choice of the static gauge manifestly 
has the $SO(6)$ $R$-symmetry since the index $\hat{I}$ is flat. 

The low energy action for the scalar fields $\varphi^{\hat{I}}$ can be obtained by expanding the bosonic part of the DBI action:
\eqn\zerotwofifteen{S_{DBI}=-\tau_3 \int d^4 x e^{-\Phi}\sqrt{-G}.}
 $G$ is the determinant of the metric
\eqn\zerotwosixteen{
G_{\mu \nu}=g_{\mu \nu} +G_{IJ}\partial_{\mu}\delta x^I \partial_{\nu}\delta x^J,}
where $g_{\mu \nu}$ is the induced metric~\zerotwoeleven\
and $G_{IJ}$ is the metric in the transverse space \zerotwofive
\eqn\zerotwoforteen{G_{IJ}=H^{-1/2} \delta_{IJ}=e^{\Phi/2}\delta_{IJ},}
where the last equality is a property of the $D7$-brane background.

Therefore,  we find that the
quadratic action for the scalar fields  in the SYM multiplet is given by
\eqn\zerotwoseventeen{
S_{Sc}=-{T_3\over 2}\hskip-3pt \int\hskip-1pt  d^4x\hskip-2pt \sqrt{-g} e^{-\Phi} 
G_{IJ} \partial_{\mu}\delta x^I \partial^{\mu}\delta x^J\hskip-2pt=
-{T_3\over 2}\hskip-3pt  \int \hskip-1pt d^4x\hskip-2pt \sqrt{-g} e^{-\Phi} 
G_{IJ} \partial_{\mu}(e^I_{\hat{I}}\varphi^{\hat{I}}) 
\partial^{\mu}(e^{J}_{\hat{J}} \varphi^{\hat{J}}),}
where the worldvolume indices $\mu$ are contracted with the induced metric \zerotwoeleven\ and we have 
used~\zerotwoforteenpointone\ to eliminate $\delta x^I$ in terms of $\varphi^{\hat{I}}$.
We note that $\partial_{\mu}$ in  \zerotwoseventeen\ acts not only on
$\varphi^{\hat{I}}$ but also on the vielbein's $e^{I}_{\hat{I}}$.
This fact  is responsible for giving a mass to the scalar fields $\varphi^{\hat{I}}$. More precisely, 
evaluating~\zerotwoseventeen\ gives
\eqn\zerotwoeighteen{
S_{Sc}=-{T_3\over 2} \int d^4x \sqrt{-g} e^{-\Phi} 
(\partial_{\mu}\varphi^{\hat{I}}\partial^{\mu}\varphi^{\hat{I}}
+{1\over 2}({\cal R}+\partial^{\mu}\partial_{\mu}\Phi)\varphi^{\hat{I}}\varphi^{\hat{I}}),}
where ${\cal R}$ is the scalar curvature of the induced metric~\zerotwoeleven, which in terms of the dilaton field $\Phi$ is given by:
\eqn\zerotwonineteen{{\cal R}=-{3 \over 8}\partial^{\mu}\Phi \partial_{\mu}\Phi-
{1 \over 2} \partial^{\mu}\partial_{\mu}\Phi.}
A similar mass term proportional to the curvature appears in the action 
of ${\cal N}=4$ Yang-Mills theory in ${\Bbb R}\times S^3$~\Nicolai\ (for a recent discussion see \Okuyama).

For later convenience, we parametrize the six scalars $\varphi^{\hat{I}}$ by a two-index  
antisymmetric tensor $\varphi^{ij}$ of $SU(4)$ via
\eqn\zerotwotwenty{
\varphi^{\hat{I}}={1\over 2}\gamma^{\hat{I}}_{ij}\varphi^{ij}, \quad
\varphi^{ij}={1\over 2}\tilde{\gamma}^{\hat{I} ij}\varphi^{\hat{I}}, \quad
\varphi_{ij}={1\over 2}\epsilon_{ijkl}\varphi^{kl},}
where $\gamma^{\hat{I}}_{ij}$ are the Clebsch-Gordan coefficients
that couple the ${\bf 6}$ representation of $SO(6)$ to the  ${\bf 4}$'s of $SU(4)$ labeled by the $i, j$ indices. 
The Clebsch-Gordan coefficients satisfy a Clifford algebra:
\eqn\zerotwotwentyone{
\{ \gamma^{\hat{I}}, \tilde{\gamma}^{\hat{J}} \} =2 \delta^{\hat{I} \hat{J}}.}
In this parametrization  the action of the scalar fields in the SYM multiplet is given by:
\eqn\zerotwotwentyonepointone{
S_{Sc}=
-{T_3\over 2} \int d^4x \sqrt{-g} e^{-\Phi} 
(\partial_{\mu}\varphi^{ij}\partial^{\mu}\varphi_{ij}
+{1\over 2}({\cal R}+\partial^{\mu}\partial_{\mu}\Phi)\varphi^{ij}\varphi_{ij}).}

Now we move on to the action for the fermions in the SYM multiplet.
As indicated earlier, 
the $\kappa$-supersymmetric DBI action depends on
thirty two spinors, which can be parametrized by two ten dimensional Majorana 
spinors of positive chirality, denoted by $\theta_1$ and $\theta_2$. Fixing $\kappa$-supersymmetry is 
equivalent to setting one of them, say $\theta_2$ to zero. Hence, the fermionic 
action can  be written in terms of $\theta_1$, which is identified with the gaugino  in the SYM multiplet. The quadratic fermionic action with fixed 
$\kappa$-supersymmetry was found in~\VanProeyentwo. Adopting their answer 
to our present case we obtain:
\eqn\zerotwotwentytwo{
S_F={T_3 \over 2} \int d^4 x \sqrt{-g} e^{-\Phi}
({\bar \theta}_1 \Gamma^{\mu} D_{\mu}\theta_1 -
{\bar \theta}_1 \hat{\Gamma}^{-1}_{D_3}(\Gamma^{\mu}W_{\mu}-\Delta)\theta_1).}
In this expression we have used:
\eqn\zerotwotwentythree{\eqalign{
&{\bar \theta}_1 =i \theta_1^{T} \gamma^{{0}}, \cr 
&\hat{\Gamma}_{D_3}=\gamma_{{0}}\gamma_{{1}}\gamma_{{2}}\gamma_{{3}}, \cr
&W_{\mu} ={1 \over 8} e^{-\Phi}\partial_{\nu}C \Gamma^{\nu}\Gamma_{\mu}, \cr
&\Delta=-{1\over 2} e^{-\Phi} \partial_{\mu}C \Gamma^{\mu}.}}

In order to write the action in terms of four dimensional spinors we use the basis of $\Gamma$ matrices in \gammamatrix\ and decompose 
\eqn\zerotwotsentyfour{
\theta_1=\pmatrix{\lambda_\alpha{}^i\cr
 {\bar \lambda}_{\dot \alpha{}i}},}
where $\lambda_{\alpha}{}^i$ is the four dimensional gaugino. We then  obtain the   action for the fermionic components of the SYM multiplet:
\eqn\zerotwotwentyfive{
S_F=T_3 \int d^4x \sqrt{-g} e^{-\Phi} 
({i \over 2} {\bar \lambda}_i\tilde{\sigma}^{\mu} D_{\mu}\lambda^i -
{i \over 2} D_{\mu}{\bar \lambda}_i\tilde{\sigma}^{\mu}\lambda^i )-
{T_3\over 4} \int d^4x \sqrt{-g} \partial_{\mu}C {\bar \lambda}_i\tilde{\sigma}^{\mu}\lambda^i .}

In summary, the total action for the SYM multiplet in the Abelian case is then given by:
\eqn\zerotwotwentysix{
S_{abel}=S_{V}+S_{Sc}+S_F,}
where $S_{V}, S_{Sc}$ and $S_F$ are given by 
\zerotwothirteen, \zerotwotwentyonepointone\ and~\zerotwotwentyfive\
respectively.

The supersymmetry transformations can be obtained from the   
superspace supersymmetry transformations on the physical fields with a compensating  $\kappa$ and diffeomorphism transformation to preserve the gauge fixing condition \VanProeyentwo. For the case under consideration we find that the action \zerotwotwentysix\ is supersymmetric under the following transformations
\eqn\zerotwotwentyseven{\eqalign{
&\delta A_{\mu}=-
i {\bar \lambda}_{i} \tilde{\sigma}_{\mu}\epsilon^i
+ {\rm c. c.} \cr
& \delta \varphi^{ij} =(\lambda^{\alpha i} \epsilon_{\alpha}{}^j
-\lambda^{\alpha j} \epsilon_{\alpha}{}^i) +
\epsilon^{ijkl}{\bar \epsilon}_{\dot\alpha k} {\bar \lambda}^{\dot\alpha}{}_l \cr
& \delta\lambda_{\alpha}{}^i =-{1 \over 2} 
F_{\mu \nu} (\sigma^{\mu} \tilde{\sigma}^{\nu})_{\alpha}^{{\ }\beta} \epsilon_{\beta}{}^i
-2 i \sigma^{\mu}_{{\ } \alpha \dot\alpha} (\partial_{\mu}\varphi^{i j})
{\bar \epsilon}^{\dot\alpha}{}_j +{i \over 2}  \sigma^{\mu}_{{\ } \alpha \dot\alpha}
(\partial_{\mu}\Phi) \varphi^{i j} {\bar \epsilon}^{\dot \alpha}{}_j,}}
where $\epsilon_{\alpha}{}^i$  is a Killing spinor satisfying \zerotwotwelve\ and subject to the constraints
\eqn\summaryconst{
 \tilde{\sigma}^{{\ }\dot\alpha \alpha}_{\bar z}\epsilon_{\alpha}{}^i=0, \quad\qquad
\tilde{\sigma}^{{\ }\dot\alpha \alpha}_{+}\epsilon_{\alpha}{}^i=0,}
so that the action is invariant under eight real supersymmetries.

We note that the variation of the gaugino contains a term proportional 
to the derivative of the dilaton which is absent in the usual 
${\cal N}=4$ SYM theory in flat space. The appearance of this term is
consistent with the presence of a  scalar ``mass term'' in the action \zerotwotwentyonepointone. The existence 
of the mass term in the action indicates that a non-vanishing constant values
of $\varphi^{ij}$ does not solve equations of motion. On the other hand, the set 
of supersymmetric solutions can be obtained by setting the variations of the fermions 
to zero. Therefore, the absence of the last term in $\delta \lambda_{\alpha}{}^i$ 
would indicate that any constant $\varphi^{ij}$ was a supersymmetric solution, 
in direct contradiction with the equations of motion\foot{One can perform  a field redefinition and get rid of the ``mass term" for the scalar fields in \zerotwotwentyone. To do this, one simply goes from  $\varphi^{\hat{I}}$ to $\delta x ^I$
\eqn\zerotwothirtyone{\delta x ^I=e^{I}{}_{\hat{I}}\varphi^{\hat{I}}=e^{-{\Phi\over 4}}\varphi^{\hat{I}}\delta^I_{\hat{I}}.}
This transformation eliminates the ``mass term" for the scalar fields  as well as the term ${i \over 2}  \sigma^{\mu}_{{\ } \alpha \dot\alpha}
(\partial_{\mu}\Phi) \varphi^{i j} {\bar \epsilon}^{\dot \alpha}{}_j$ in the supersymmetry transformations for the gauginos.}. 

The formalism using the covariant $D$-brane action allowed us to write a 
supersymmetric gauge theory action when the gauge group is Abelian. 
We now extend the analysis of the action and the supersymmetry transformations 
to the case when the gauge group is non-Abelian. The extension is relatively 
straightforward. In the action \zerotwotwentysix\ we replace all derivatives 
$D_\mu$ by the gauge covariant derivatives ${\cal D}_\mu$, where
\eqn\gaugecov{
{\cal D}_\mu\cdot=D_\mu\cdot+[A_\mu,\cdot],}
replace the Chern-Simons term in~\zerotwothirteen\ 
by its non-Abelian analog
\eqn\zerotwothirteennew{
-{T_3 \over 4}\int d^4 x \sqrt{-g}\partial_{\mu}C 
\epsilon^{\mu \nu \rho \sigma}\hbox{Tr}
(A_{\nu} F_{\rho \sigma}-{2 \over 3}A_{\nu} A_{\rho} A_{\sigma}),}
and add the familiar non-Abelian couplings of ${\cal N}=4$ SYM in flat space:
\eqn\zerotwentyeight{S_{nabe}=T_3\int d^4x \sqrt{-g} e^{-\Phi}\hbox{Tr}({\bar\lambda}_{{\dot \alpha}i}[{\bar\lambda}^{\dot \alpha}{}_{j},\varphi^{ij}]+  \lambda^{\alpha i}[{\lambda}_{\alpha}{}^{j},\varphi_{ij}] -{1\over 2}[\varphi^{ij},\varphi^{kl}][\varphi_{ij},\varphi_{kl}]).}

In the supersymmetry transformations \zerotwotwentyseven\ we replace also all covariant derivatives $D_\mu$ with  ${\cal D}_\mu$, and add to $\delta\lambda_{\alpha}{}^i$  the usual flat space ${\cal N}=4 $SYM commutator term $-2[\varphi_{jk},\varphi^{ki}]\epsilon_{\alpha}{}^j$. 

We have found the complete non-Abelian action on $N$ $D3$-branes when embedded in the $D7$-brane background. The full action is given by:
\eqn\zerotwotwentynine{\eqalign{
S=&-{T_3 \over 4}\int d^4 x \sqrt{-g} e^{-\Phi}\hbox{Tr}F_{\mu \nu}F^{\mu \nu}
-{T_3 \over 4}\int d^4 x \sqrt{-g} \partial_{\mu}C 
\epsilon^{\mu \nu \rho \sigma}\hbox{Tr}\left(A_{\nu} F_{\rho \sigma}-
{2 \over 3} A_{\nu} A_{\rho} A_{\sigma}\right)\cr
&+T_3 \int d^4x \sqrt{-g} e^{-\Phi} 
\hbox{Tr}\left({i \over 2} {\bar \lambda}_{i}\tilde{\sigma}^{\mu} {\cal D}_{\mu}\lambda^i -
{i \over 2} {\cal D}_{\mu}{\bar \lambda}_{i}\tilde{\sigma}^{\mu}\lambda^i\right)-
{T_3\over 4} \int d^4x \sqrt{-g} \partial_{\mu}C \hbox{Tr} ({\bar \lambda}_{i}\tilde{\sigma}^{\mu}\lambda^i)\cr &-{T_3\over 2} \int d^4x \sqrt{-g} e^{-\Phi} 
\hbox{Tr}\left({\cal D}_{\mu}\varphi_{ij}{\cal D}^{\mu}\varphi^{ij}
+{1\over 2}({\cal R}+\partial^{\mu}\partial_{\mu}\Phi)\varphi^{ij}\varphi_{ij}\right)\cr
&+T_3\int d^4x \sqrt{-g} e^{-\Phi}\hbox{Tr}\left({\bar\lambda}_{{\dot \alpha}i}[{\bar\lambda}^{\dot \alpha}{}_{j},\varphi^{ij}]+  \lambda^{\alpha i}[{\lambda}_{\alpha}{}^{j},\varphi_{ij}] -{1\over 2}[\varphi^{ij},\varphi^{kl}][\varphi_{ij},\varphi_{kl}]\right). }}

The action on the $D3$-branes \zerotwotwentynine\ is invariant under the following explicit supersymmetry transformations 
\eqn\zerotwothirty{\eqalign{
\delta A_{\mu}=&-
i {\bar \lambda}_{i} \tilde{\sigma}_{\mu}\epsilon^{i}
+ {\rm c. c.} \cr
 \delta \varphi^{ij} =&(\lambda^{\alpha i} \epsilon_{\alpha}{}^j
-\lambda^{\alpha j} \epsilon_{\alpha}{}^i) +
\epsilon^{ijkl}{\bar \epsilon}_{\dot\alpha k} {\bar \lambda}^{\dot\alpha}{}_l \cr
 \delta\lambda^i_{\alpha} =&-{1 \over 2} 
F_{\mu \nu} (\sigma^{\mu} \tilde{\sigma}^{\nu})_{\alpha}^{{\ }\beta} \epsilon_{\beta}{}^i
-2 i \sigma^{\mu}_{{\ } \alpha \dot\alpha} (\partial_{\mu}\varphi^{i j})
{\bar \epsilon}^{\dot\alpha}{}_j +{i \over 2}  \sigma^{\mu}_{{\ } \alpha \dot\alpha}
(\partial_{\mu}\Phi) \varphi^{i j} {\bar \epsilon}^{\dot \alpha}{}_j\cr
&-2[\varphi_{jk},\varphi^{ki}]\epsilon_{\alpha}{}^j,}}
where  $\epsilon_{\alpha}{}^i$  is a Killing spinor satisfying \zerotwotwelve\ and subject to the constraints
\summaryconst. It, thus, preserves the same eight supersymmetries
preserved by the $D$-brane intersection.

The detailed check of  the invariance of the action \zerotwotwentynine\  under the supersymmetry transformations \zerotwothirty\ is summarized in Appendix B. 

We finish this subsection by stating the symmetries of this field theory. 
The bosonic symmetry is $ISO(1, 1)\times SO(6)$. Furthermore, the 
field theory is invariant under eight real supercharges. Note that the 
theory is not conformally invariant. The dilatations and special conformal 
transformations are broken by $z$-dependent warp-factors $H_7(z, \bar z)$ 
in~\zerotwoeleven.



\subsec{The WZW Surface Operator}


In this final subsection we show that the field theory on the $D3$/$D7$ intersection describes a surface  operator of  ${\cal N}=4$ SYM in the $D7$-brane background.
 This surface operator, unlike the one in \GukovJK, has a classical expression that can be written down in terms of the classical fields that appear in the Lagrangian of ${\cal N}=4$ SYM. 
 
The strategy that we follow for determining the expression for the surface operator is to integrate out explicitly the  fermions $\chi, {\bar \chi}$ that are localized on the surface. The effect of the non-dynamical $D7$-brane gauge field is trivial and we suppress it in this section. In Appendix $C$ we show that integrating over this gauge field reproduces the same answer as when we suppress it.
 This same strategy was used in \GomisSB\  to derive the Wilson loop operators in ${\cal N}=4$, which were obtained  by integrating out the localized degrees of freedom living on the loop arising from a brane intersection.
 
We want to perform the following path integral\foot{After this integral is performed, we must still integrate over the ${\cal N}=4$ SYM fields.}
\eqn\path{
Z=e^{iS}\cdot\int [D\chi][D {\bar \chi}] \exp\left(iS_{defect}\right),}
where:
\eqn\fermions{
S_{defect}=\int dx^+dx^- {\bar \chi} \left(\partial_++A_+\right)\chi.}
$S$ is the ${\cal N}=4$ SYM action in the $D7$-brane background
\zerotwotwentynine.

We proceed to integrating out the chiral fermions localized on the surface. This is well known to produce a WZW model, which precisely captures the anomaly of the chiral fermions via the identity
\eqn\dirac{
\hbox{Det}(\partial_++A_+)=\exp{\left(ic_R \Gamma_{WZW}(A)\right)},}
where $c_R$ is the index of the representation $R$ under which the fermions transform. The explicit expression for the WZW action one gets is
\eqn\WZW{\eqalign{
\Gamma_{WZW}(A)&=-{1\over 8\pi}\int dx^+dx^- \hbox{Tr}\left[ \left(U^{-1}\partial_+ U\right)\left(U^{-1}\partial_- U\right)-\left(U^{-1}\partial_+ U\right)\left(V^{-1}\partial_- V\right)\right]\cr
&-{1\over 24 \pi}\int d^3x \epsilon^{ijk}\hbox{Tr}\left[\left(U^{-1}\partial_i U\right)\left(U^{-1}\partial_j U\right)\left(U^{-1}\partial_k U\right)\right],}}
 where $U$ and $V$ are $U(N)$ group elements nonlocally related to the gauge field of ${\cal N}=4$ SYM:
 \eqn\nonlocal{
 A_{+}=U^{-1}\partial_+ U\qquad \qquad  A_{-}=V^{-1}\partial_- V.}
 We note that $\Gamma_{WZW}(A)$ differs from the conventional WZW model action by the 
addition of a local counterterm:
\eqn\counter{
{1\over 8\pi}\int dx^+dx^- \hbox{Tr}
\left[\left(U^{-1}\partial_+ U\right)\left(V^{-1}\partial_- V\right)\right]
={1 \over 8}\int d x^{+} d x^{-}\hbox{Tr} A_{+} A_{-}.}
The addition of this term is needed to guarantee that \WZW\ reproduces the correct chiral anomaly. Indeed, it is straightforward to show that under a $U(N)$ gauge transformation $\delta A_\mu=\partial_\mu+[A_\mu ,L]$ we have that the WZW action \WZW\ is not invariant:
\eqn\anomaa{
\delta \Gamma_{WZW}(A)={1\over 8\pi}\int dx^+ dx^- \hbox{Tr}\left[L\left(\partial_+A_- - \partial_-A_+\right)\right].}
This gives  the same anomalous variation as the
usual  anomaly in two dimensions \aly. We recall that   our complete action, which combines the ${\cal N}=4$ SYM action on the $D7$-brane background \zerotwotwentynine\ with the defect term in \fermions\ is not anomalous.  
The anomaly produced by the WZW action is precisely cancelled by a Chern-Simons term.

We also note that $\Gamma_{WZW}(A)$ is not invariant under the supersymmetry transformations \zerotwothirty, unlike the original action $S_{defect}$.  But we recall that the Chern-Simons terms always has a boundary term under any variation of the gauge field and that this boundary contribution cancels the variation of $\Gamma_{WZW}(A)$ proportional to $\delta A_-$. For this cancellation to occur, it is also crucial to add the local counterterm  \counter.

 Therefore, integrating out the localized fields has the effect of inserting the following surface operator into the  gauge theory action \zerotwotwentynine:
 \eqn\insertop{
{\cal O}_\Sigma=\exp\left(iM\Gamma_{WZW}(A)\right).}
The surface operator is described by a $U(N)$ WZW model at level $M$.
The explicit form of the action is  \WZW, where $U$ and $V$ are $U(N)$ valued group elements that are nonlocally related to the ${\cal N}=4$ SYM gauge fields via:
\eqn\relatte{
A_+=U^{-1}\partial_+U\qquad A_-=V^{-1}\partial_-V.}
The surface operator \insertop\ is  supersymmetric under the transformations  \zerotwothirty\ and $U(N)$ invariant when combined with the gauge theory action in the $D7$-brane background \zerotwotwentynine.

Using  the explicit expression for the surface operator one can study its properties in 
perturbation theory. For the case when $\Sigma={\Bbb R}^{1,1}$ we expect that 
supersymmetry requires $\la {\cal O}_\Sigma\ra=1$, just like in the case of the Wilson line.  
Another interesting case to consider -- which is related by a conformal 
transformation to the Euclidean version of the previous case --  is when 
$\Sigma=S^2$. In this case   $\Sigma$ is curved and we 
expect that there is a conformal anomaly associated with the surface 
which would be interesting to compute explicitly. The bulk description 
discussed in the next section supports these expectations, 
as we find that at least in the probe 
approximation that $\la {\cal O}_\Sigma\ra=1$ and that 
there is a conformal anomaly for the cases $\Sigma={\Bbb R}^{1,1}$ and $S^2$ respectively.

Given that these operators are supersymmetric  one may expect that the computation of their expectation value is captured by a simpler model, similar to what happens for circular Wilson 
loops~\EricksonAF, \DrukkerRR.
One may be able to derive  the reduced model by topologically twisting the gauge theory by the supercharges preserved by the surface operator.

\newsec{The Bulk Description}

In this section we study the physics of the surface operator from a dual gravitational point of view. We find that there is a regime in which the $D7$-branes can be treated as a probe brane in 
$AdS_5\times S^5$ and identify the corresponding regime in the gauge theory. 
We also find the exact solutions of the Type IIB supergravity equations of motion -- which  
take the backreaction 
of the $D7$-branes into account -- which are dual to the 
surface operators in the gauge  theory we have constructed in this paper. 

\subsec{The Probe Approximation and Anomaly Suppression}

In the previous section we have constructed the decoupled low energy effective field theory living on the $D3$/$D7$ intersection \braneconf. 
Following 
\lref\MaldacenaRE{
  J.~M.~Maldacena,
  ``The large N limit of superconformal field theories and supergravity,''
  Adv.\ Theor.\ Math.\ Phys.\  {\bf 2}, 231 (1998)
  [Int.\ J.\ Theor.\ Phys.\  {\bf 38}, 1113 (1999)]
  [arXiv:hep-th/9711200].
}
\MaldacenaRE\ our aim in this section is to provide the bulk gravitational description of this field theory. This requires finding the 
supergravity solution describing the brane intersection \braneconf\ 
\MaldacenaRE.

In the absence of the $D7$-branes, the gauge theory on $N$ $D3$-branes is dual to string theory in 
$AdS_5 \times S^5$ \MaldacenaRE. We are interested in understanding 
what the effect of introducing the $D7$-branes is in the bulk description. 

One may try to first consider the $D7$-branes as a small perturbation 
around the $AdS_5 \times S^5$ background. 
The parameter that controls the gravitational backreaction due to the $M$ $D7$-branes  can be extracted from the supergravity equations of motion. It is governed by 
\eqn\smallpara{
\epsilon=M\cdot G_{10}\tau_7=g_s M = {g^2\over 2\pi}  M.}
In the last step we have written the parameter using gauge theory variables, 
where $g$ is the gauge theory coupling constant. 
Therefore, we can treat the $D7$-branes as probes in $AdS_5\times S^5$ as long as $g^2 M$ is small. 

In the regime where $g^2 M$ is small we can consistently treat the $D7$-branes 
in the probe approximation. It is straightforward to show that the $D7$-brane equations 
of motion are solved by the embedding \braneconf\ even when we place the $D7$-branes in 
the non-trivial supergravity background produced by the $D3$-branes. 
Upon taking the $D3$-brane near horizon limit, the brane embedding geometry is that of 
$AdS_3\times S^5$ 
\lref\SkenderisVF{
  K.~Skenderis and M.~Taylor,
  ``Branes in AdS and pp-wave spacetimes,''
  JHEP {\bf 0206}, 025 (2002)
  [arXiv:hep-th/0204054].
}
\SkenderisVF.

We are now in a position to determine what is the field theory counterpart 
of the bulk probe approximation. We recall that the gauge theory we constructed 
in the previous section is defined on the $D7$-brane background. 
In the probe regime, where $g^2M<<1$, the background produced by the $D7$-branes 
becomes trivial, as the metric becomes flat, the dilaton goes to a 
constant and the RR flux vanishes. Hence, in this limit we get the following gauge theory
\eqn\limitgauge{
S=S_{{\cal N}=4}+\int dx^+dx^ -\ {\bar \chi}(\partial_++A_+)\chi,}
where $S_{{\cal N}=4}$ is the standard action of ${\cal N}=4$ SYM in flat space.

However, we have argued that it was crucial to 
consider the gauge theory on the full $D7$-brane geometry, so as to get an 
anomaly free and supersymmetric theory. The resolution lies in the observation that the gauge anomaly is suppressed in this limit. In order to better understand the parameter controlling the anomaly, it is convenient to rescale the gauge fields in the action as follows $A_{\mu}\rightarrow g A_{\mu}$. In this presentation it becomes clear what the effect of the coupling constant is on physical quantities. The quantum effective action obtained by integrating the fermions is anomalous, the obstruction to gauge invariance being measured by\foot{In the frame where the coupling constant controls the interaction vertices in gauge theory, the gauge parameter  must also be rescaled  $L\rightarrow gL$.}
 \eqn\alya{\delta_{L}S={g^2M\over 8 \pi}\int d x^{+} d x^{-}\hbox{Tr}_{U(N)}(LdA),}
so that the anomaly is controlled by the same parameter that 
controls the backreaction of the $D7$-branes in the bulk \smallpara, 
and is therefore suppressed in the probe limit $g^2M\rightarrow 0$.

Note that 
to leading order in the $g^2M$ expansion the two dimensional Poincare symmetry of the 
gauge theory is enlarged to $SO(2,2)\simeq 
SL(2,{\Bbb R})\times SL(2, {\Bbb R})$, 
as long as the $D7$-branes are coincident.  This can be understood from 
the point of view of the symmetries of ${\cal N}=4$ SYM in flat space. 
A surface $\Sigma={\Bbb R}^{1,1}\subset {\Bbb R}^{1,3}$ is invariant under 
an   $SO(2,2)$ 
subgroup of the $SO(2,4)$ four dimensional conformal group. 
The symmetries are generated by $P_{\mu}, M_{\mu\nu}, K_{\mu}$ and $D$, with $\mu=0,1$, 
where 
$K_\mu$ and $D$ generate the special conformal and dilatation 
transformations respectively. In this case -- where the $D7$-branes are coincident -- the 
theory acquires eight extra supersymmetries, which correspond to 
conformal supersymmetries. 
Indeed, the$S_{{\cal N}=4}$ term
 in~\limitgauge\ is invariant 
under sixteen superconformal supersymmetries generated by $\varepsilon_{\alpha}^{{\ }i}$.
The second term in~\limitgauge, given by $S_{defect}$,   is invariant under the  conformal  supersymmetries generated by:
\eqn\limitgaugeone{
\tilde{\sigma}_{-}^{{\ }\dot\alpha \alpha}\varepsilon_{\alpha}^{{\ }i}=0.}
To see this consider the relevant superconformal transformations
\eqn\limitgaugetwo{ 
\delta A_{\mu}=-i x^{\nu} \lambda^{\beta i} 
\sigma_{\mu \beta \dot\alpha}
\tilde{\sigma}_{\nu}^{{\ }\dot\alpha \alpha}\varepsilon_{\alpha}^{{\ }i} +{\rm c.c.}, 
\qquad \delta{\chi}=0.}
It is straightforward to show that $\delta A_+=0$ 
if~\limitgaugeone\ is fulfilled and the defect action is localized at $z=0$.
All these symmetries combine into the $SU(1,1|4)\times SL(2, {\Bbb R})$ supergroup
\lref\VanProeyenNI{
  A.~Van Proeyen,
  ``Tools for supersymmetry,''
  arXiv:hep-th/9910030.
}
\VanProeyenNI.

Once $g^2M$ corrections are taken into account, so that the anomaly, 
the Chern-Simons terms and the $D7$-brane background cannot be neglected,  
the symmetries are broken down\foot{This is similar to   the breaking of conformal 
invariance by $g^2M$ effects that occurs when considering ${\cal N}=4$ SYM 
coupled to $M$ hypermultiplets, where the $\beta$-function is 
proportional to $g^2M$, so that $g^2M$ effects break conformal symmetry.}
to $ISO(1,1)\times SO(6)$ and the theory is invariant 
under eight supersymmetries. Even if we start with coincident 
$D7$-branes, once one takes into account the proper global 
solution \zerotwoten, the $U(1)$ symmetry is broken.

Let's now consider the symmetries of the bulk theory in the probe approximation. 
When the $M$ $D7$-branes are coincident the $D7$-branes are invariant under 
$SO(2,2)\times SO(2)\times SO(6)$. The $SO(2,2)$  and $SO(6)$ symmetries act  
by isometries on the $AdS_3$ and $S^5$ worldvolume geometry respectively. 
The $U(1)$ symmetry rotates the $z$-plane in \braneconf. 
We  show (see Appendix D) that the coincident 
$D7$-branes also preserve half of the Type IIB supersymmetries, which 
coincide precisely with the Poincare and special conformal supersymmetries 
preserved in the gauge theory, which are given by 
\eqn\poinsusy{
\tilde{\sigma}_+{}^{\dot{\alpha}\alpha}\epsilon_{\alpha}{}^i=0}
and  
\eqn\confsusy{
\tilde{\sigma}_-{}^{\dot{\alpha}\alpha}\varepsilon_{\alpha}{}^i=0}
respectively. The unbroken symmetries combine to form a chiral 
superconformal group, which is an $SU(1,1|4)\times SL(2, {\Bbb R})$ supergroup, 
as thus coincides with the gauge theory symmetries discussed above. 
If the $D7$-branes are not coincident, 
in both field and gravity theory
the symmetry is broken down to $ISO(1,1)\times SO(6)$ 
and only eight supersymmetries survive.

The $AdS_3\times S^5$ $D7$-brane ends on the surface $\Sigma$ on the boundary of 
$AdS_5 \times S^5$, 
thus providing boundary conditions for the surface operator. One can use the probe $D7$-brane to calculate the expectation value of the surface operator in the probe regime. In the semiclassical approximation it  is given by
\lref\WittenQJ{
  E.~Witten,
  ``Anti-de Sitter space and holography,''
  Adv.\ Theor.\ Math.\ Phys.\  {\bf 2}, 253 (1998)
  [arXiv:hep-th/9802150].
}
\lref\GubserBC{
  S.~S.~Gubser, I.~R.~Klebanov and A.~M.~Polyakov,
  ``Gauge theory correlators from non-critical string theory,''
  Phys.\ Lett.\  B {\bf 428}, 105 (1998)
  [arXiv:hep-th/9802109].
}
\WittenQJ\GubserBC\ 
\eqn\expectcalc{
\la {\cal O}_\Sigma \ra=\exp(-S^{on-shell}_{D7}).}
For the brane embedding at hand the $D7$-brane on-shell action is given by
\eqn\actiononshell{
S^{on-shell}_{D7}=\tau_7L^8\hbox{vol}(S^5)\hbox{vol}_{ren}(AdS_3),}
where $L$ is the $AdS_5/S^5$ radius, $\hbox{vol}(S^5)$ is the volume of the $S^5$ and 
$\hbox{vol}_{ren}(AdS_3)$ is the renormalized volume of $AdS_3$. As usual the bulk action is infrared divergent and requires renormalization. This is accomplished by adding covariant counterterms. It is easy to show that the renormalized volume of $AdS_3$  
vanishes so we find that $\la {\cal O}_\Sigma \ra=1$ in the probe approximation. The same answer is obtained for the gauge theory in the probe approximation \limitgauge, as one just gets the partition function over free fermions. 

One may consider surface operators defined on surfaces $\Sigma$ other than 
${\Bbb R}^{1,1}$ in the probe approximation.  In the bulk, this corresponds to considering $D7$-brane solutions of the DBI equations of motion that end on the boundary of $AdS_5 \times S^5$ on $\Sigma$. 
The case when $\Sigma=S^2$ can be obtained easily from the euclidean solution 
with $\Sigma={\Bbb R}^2$ by acting with a broken special conformal transformation. 
In this case the bulk $D7$ brane is still $AdS_3 \times S^5$, but now 
$AdS_3$ is in global coordinates and the brane ends on the boundary of 
$AdS_5$ on an $S^2$. In this case, the calculation of $D$-brane action is non-trivial as the 
renormalized volume of global $AdS_3$ is non-trivial. In this case, one finds 
that the $D7$-brane has a conformal anomaly, similar to the one discussed in 
\lref\GrahamPM{
  C.~R.~Graham and E.~Witten,
  ``Conformal anomaly of submanifold observables in AdS/CFT correspondence,''
  Nucl.\ Phys.\  B {\bf 546}, 52 (1999)
  [arXiv:hep-th/9901021].
}
\lref\BerensteinIJ{
  D.~E.~Berenstein, R.~Corrado, W.~Fischler and J.~M.~Maldacena,
  ``The operator product expansion for Wilson loops and surfaces in the  large
  N limit,''
  Phys.\ Rev.\  D {\bf 59}, 105023 (1999)
  [arXiv:hep-th/9809188].
}
\BerensteinIJ, \GrahamPM\ in the context of $M2$-branes ending on an $S^2$ in 
$AdS_7\times S^4$. 
This is encoded in the coefficient of the logarithmic divergence of the on-shell action, 
which for M $D7$-branes is controlled by $\tau_7 L^8=g^2 M N^2$. It would be interesting to calculate the corresponding conformal anomaly in the gauge theory in perturbation theory.

\subsec{The Supergravity Solution}

In this final subsection we find the bulk description of a surface operator in terms of solution of the
Type IIB supergravity. 
According to AdS/CFT duality, this solution is obtained by taking 
the near-horizon limit of the supergravity solution of the brane intersection~\braneconf.
The explicit form of the solution corresponding to~\braneconf\
is given by:
\eqn\solusugra{\eqalign{
ds^2&=-H_3^{-1/2}H_7^{-1/2}dx^+dx^-+
H_3^{-1/2}H_7^{1/2}dzd\bar{z}+
H_3^{1/2}H_7^{-1/2}dx^Idx^I\cr
e^{-\Phi}&=H_7\cr
F_{0123I}&=H_7\partial_IH_3^{-1}\cr
\partial_{\bar{z}}\tau&=0\qquad \hbox{where}\qquad \tau=C+ie^{-\Phi}.}}
$H_3=H_3(x^I)$ is an arbitrary harmonic function in the space transverse to the $D3$-branes while $H_7=H_7(z,\bar{z})$ determines the $D7$-brane contribution and it is of the same form as in section $2$. It is straightforward to show that this supergravity background solves the Type IIB supergravity Killing spinor equations and that moreover the space of solutions is eight real dimensional and can be parametrized by four dimensional spinors satisfying the constraints \susy\ and \susyz. 

Here we are interested in the supergravity solution describing the decoupled gauge theory constructed earlier and that lives on the $D3/D7$ intersection. This corresponds to taking the near horizon limit of the supergravity solution corresponding to the case when the $N$ $D3$-branes are coincident -- so that  $H=1+L^4/\rho^4$ -- where $dx^Idx^I=d\rho^2+\rho^2d\Omega_5$. In this limit the metric can be written in 
terms of an $AdS_3 \times S^5$ factor. The geometry describing the surface operator is given by
\eqn\geomsurf{
ds^2=H_7^{-1/2}\left(ds^2_{AdS_3}+L^2 d\Omega_5\right)+{\rho^2\over L^2}H_7^{1/2}dzd{\bar z},}
where:
\eqn\adsmetric{
ds^2_{AdS_3}=-{\rho^2\over L^2}dx^+dx^-+L^2{d\rho^2\over \rho^2}.}

This metric reveals several interesting features of the holographic correspondence. We have argued that it is inconsistent to describe a low energy field theory with anomaly inflow by treating the gauge theory in flat space. We have argued that the proper description of the system is  in terms of the gauge theory in the supergravity background produced by the other brane. In particular, for our intersection, we have constructed the gauge theory on the $D3$-branes in the background of the $D7$-branes and found that this field theory has all the expected properties. We can now use the dual supergravity solution \adsmetric\ to indeed infer that the holographic dual gauge theory lives in the background geometry of the $D7$-branes and not in flat space. Indeed if we analyze the metric living on the conformal boundary -- where $\rho\rightarrow \infty$ -- we precisely get the metric on which the gauge theory lives \zerotwoeleven. 

The solution also gives information about the non-perturbative behavior of the 
symmetries of the gauge theory. As we discussed earlier in this section,
the gauge theory has $SO(2,2)$ symmetry to leading order in a $g^2M$ expansion. 
This symmetry is intimately related to the geometrical surface on which the fermions live. 
However, once the $g^2 M$ corrections are turned on and the $D7$-brane 
backreaction cannot neglected, the conformal symmetry is broken. 
The dual geometry~\geomsurf\ has the same symmetries. In particular, 
the $SO(2,2)$ symmetry is broken down to $ISO(1, 1)$. 
First, the warp-factor $H_7$ is not invariant under dilatations 
and special conformal transformations just like in field theory. 
Second, 
we see that the $AdS_3$ radial coordinate $\rho$ does not decouple from the transverse space 
and appears explicitly in the transverse metric. 
As usual, the $SO(2,2)$ conformal transformations correspond to $AdS_3$ isometries. However, since 
$AdS_3$ isometries act non-trivially on $\rho$ and $z$
we find that the  $SO(2,2)$ conformal symmetry of the surface operator is broken down 
to $ISO(1,1)$. 
The supersymmetries are also reduced with respect to the probe approximation. 
This can be shown (see Appendix E) by explicitly solving the Type IIB 
Killing spinor equations in the background~\geomsurf. 
The explicit Killing spinor is given by
%
\eqn\kill{
\epsilon=h(\theta,\varphi_a))H_7^{-1/2}\rho^{1/2}\epsilon_{0},}
where $h(\theta,\varphi_a)$ is the standard contribution from $S^5$~\Pope, \Claus\
(see Appendix D for the explicit expression).
In addition, $\epsilon$ (as well as $\epsilon_{0}$) is 
subject to the constraints \susy\ and~\susyz, which give rise to eight real supersymmetries. 
Thus, we obtain the same symmetries as those preserved by the gauge theory.

\bigbreak\bigskip\bigskip\centerline{{\bf Acknowledgements}}\nobreak

We thank L. Freidel, R. Myers, A. Tseytlin and specially D. Sorokin for discussions. J.G. thanks l'\'Ecole Polytechnique for hospitality and the European Excellence Grant MEXT-CT-2003-509661 for partial support. This research was supported by Perimeter Institute for Theoretical
Physics.  Research at Perimeter Institute is supported by the Government
of Canada through Industry Canada and by the Province of Ontario through
the Ministry of Research and Innovation. J.G.  also acknowledges further  support by an NSERC Discovery Grant.



\medskip\medskip\medskip\medskip


\appendix{A}{The $\sigma$-Matrix Conventions}


The $\sigma$-matrices $\sigma_{\mu}^{{\ }\alpha \dot \alpha}$  
are defined in the usual way:\foot{In this Appendix 
the index $\mu$ is assumed to be flat. In curved space-time we will have to replace 
in all expressions $\eta_{\mu \nu}$ by the space-time metric.}
\eqn\Aone{
\sigma_{0}=\left(\eqalign{&1 \qquad 0 \cr  
&0 \qquad 1 }\right),\quad 
\sigma_{1}=\left(\eqalign{&0 \qquad 1 \cr  
&1 \qquad 0 }\right), \quad
\sigma_{2}=\left(\eqalign{&0 \quad -i \cr  
&i \qquad 0 }\right), \quad
\sigma_{3}=\left(\eqalign{&1 \qquad 0 \cr  
&0 \quad -1 }\right).}
In addition, we define:
\eqn\Atwo{\eqalign{
&\tilde{\sigma}_{\mu}^{{\ }\dot\alpha \alpha}=
\epsilon^{\dot\alpha \dot\beta}\epsilon^{\alpha \beta}
\sigma_{\mu \beta \dot\beta}, \cr
&\tilde{\sigma}_{\mu}= (\sigma_{0}, -\sigma_1, -\sigma_2, -\sigma_3).}}
These matrices satisfy the following properties:
\eqn\Athree{\eqalign{
&(\sigma_{\mu}\tilde{\sigma}_{\nu}+\sigma_{\nu}\tilde{\sigma}_{\mu})_{\alpha}^{{\ }\beta}=
-2 \eta_{\mu \nu} \delta_{\alpha}^{{\ }\beta}, \cr
&(\tilde{\sigma}_{\mu}\sigma_{\nu}+\tilde{\sigma}_{\nu}\sigma_{\mu})_{{\ }\dot\beta}^{\dot\alpha}=
-2 \eta_{\mu \nu} \delta_{{\ }\dot\beta}^{\dot\alpha}, \cr
&{\rm tr}(\sigma_{\mu} \tilde{\sigma}_{\nu})=-2 \eta_{\mu \nu}, \cr
&\sigma^{\mu}_{{\ }\alpha \dot\alpha} \tilde{\sigma}_{\mu}^{{\ }\dot\beta \beta}=
-2 \delta_{\alpha}^{\beta}\delta_{\dot\alpha}^{\dot\beta}, \cr
&\sigma_{\mu}\tilde{\sigma}_{\nu}\sigma_{\rho}=
(\eta_{\mu \rho}\sigma_{\nu}-\eta_{\nu \rho} \sigma_{\mu}-
\eta_{\mu \nu}\sigma_{\rho}) +i \epsilon_{\mu \nu\rho \sigma}\sigma^{\sigma}, \cr
&\tilde{\sigma}_{\mu}\sigma_{\nu}\tilde{\sigma}_{\rho}=
(\eta_{\mu \rho}\tilde{\sigma}_{\nu}-\eta_{\nu \rho}\tilde{\sigma}_{\mu}-
\eta_{\mu \nu}\tilde{\sigma}_{\rho}) -i \epsilon_{\mu \nu\rho \sigma}\tilde{\sigma}^{\sigma}.}}
In the paper we go from coordinates $x^{\mu}$ to: 
\eqn\Afour{
x^{\pm}=x^0 \pm x^{1}, \quad z= x^2+i x^3.}
In this basis, we obtain:
\eqn\Afive{\eqalign{
&\sigma^{-}=\sigma^0 -\sigma^1 =-\sigma_0 -\sigma_1=-
\left(\eqalign{&1 \qquad 1 \cr  
&1 \qquad 1 }\right), \cr
&\tilde{\sigma}^{-}=\tilde{\sigma}^0 -\tilde{\sigma}^1=
-\tilde{\sigma}_0 -\tilde{\sigma}_1=-\sigma_0+\sigma_1=
\left(\eqalign{&-1 \qquad 1 \cr  
&\quad\quad 1 \quad -1 }\right), \cr
&{\rm etc}.}}
In particular, we have
\eqn\Asix{
\tilde{\sigma}_{+}=\eta_{+ -}\tilde{\sigma}^{-}=-{1\over 2} \tilde{\sigma}^{-}
={1\over 2} (\tilde{\sigma}_0+\tilde{\sigma}_1) ={1\over 2}
\left(\eqalign{&\quad\quad1 \quad -1 \cr  
&-1 \qquad 1 }\right)}
and:
\eqn\Aseven{
\tilde{\sigma}_{\bar z}=\eta_{{\bar z} z}\tilde{\sigma}^{z}=
{1\over 2} (\tilde{\sigma}_2+i \tilde{\sigma}_3) ={1\over 2}
\left(\eqalign{&-i \qquad i \cr  
&-i \qquad i }\right).}
The restriction on the supersymmetry parameter \four\ found in the paper can be written as:
\eqn\Aeight{
\tilde{\sigma}_{+}^{{\ }\dot\alpha \alpha}\epsilon_{\alpha}^{{\ }i}=0
\qquad {\rm or} \qquad
\tilde{\sigma}_{{\bar z}}^{{\ }\dot\alpha \alpha}\epsilon_{\alpha}^{{\ }i}=0.}
Both equations in~\Aeight\ imply that  $\epsilon_1^{{\ }i}=\epsilon_2^{{\ }i}$.


\appendix{B}{Explicit Check of the Supersymmetry of the Action}


In this Appendix, we explicitly show that the non-Abelian $D3$-brane action in the $D7$-brane background given in \zerotwotwentynine\ is invariant under the supersymmetry transformations in \zerotwothirty.
The action has the following structure
\eqn\Bone{S=S_{V}+S_{Sc}+S_{F}+S_{nab},}
where: 
\eqn\Btwo{\eqalign{
&S_V=\hskip-2pt -{T_3 \over 4}\int d^4 x \sqrt{-g} e^{-\Phi}{\rm Tr}
F_{\mu \nu}F^{\mu \nu}-
{T_3 \over 4}\int d^4 x \sqrt{-g} \partial_{\mu}C 
\epsilon^{\mu \nu \rho \sigma}{\rm Tr}
\left(A_{\nu} F_{\rho \sigma}-
{2 \over 3} A_{\nu} A_{\rho} A_{\sigma}\right), \cr 
&S_{Sc}=-{T_3\over 2} \int d^4 \sqrt{-g} e^{-\Phi}{\rm Tr} 
\left({\cal D}_{\mu}\varphi^{ij}{\cal D}^{\mu}\varphi_{ij}
+{1\over 2}({\cal R}+\partial^{\mu}\partial_{\mu}\Phi)\varphi^{ij}\varphi_{ij}\right), \cr
&S_F=T_3 \int d^4x \sqrt{-g} e^{-\Phi} {\rm Tr}
\left({i \over 2} {\bar \lambda}_i\tilde{\sigma}^{\mu} {\cal D}_{\mu}\lambda^i -
{i \over 2} {\cal D}_{\mu}{\bar \lambda}_i\tilde{\sigma}^{\mu}\lambda^i \right)-
{T_3\over 4} \int d^4x \sqrt{-g} \partial_{\mu}C {\bar \lambda}_i\tilde{\sigma}^{\mu}\lambda^i, \cr
&S_{nab}= T_3 \int d^4 x \sqrt{-g} e^{-\Phi} {\rm Tr}
\left({\bar \lambda}_{\dot\alpha i}[{\bar \lambda}^{\dot\alpha}_{{\ }j}, \varphi^{i j}]
+\lambda^{\alpha i}[\lambda_{\alpha}^{{\ }j}, \varphi_{i j}]
-{1\over 2} [\varphi^{i j}, \varphi^{k l}]
[\varphi_{i j}, \varphi_{k l}]\right).}}
In looking at the supersymmetry variation of the  action we do not write the terms that cancel 
exactly in the same way as they cancel in ${\cal N}=4$ SYM theory in flat space. 
That is,  we  only keep the terms which contain derivatives of the background  supergravity 
fields and $\epsilon^i$ and discuss how they cancel. 
Let us first look at the variation of the terms in the action involving the gauge fields  $S_V$. We obtain:
\eqn\Bthree{\eqalign{
\delta S_{V}=&-{T_3 \over 2}\int d^4 x \sqrt{-g}\, \partial_{\mu}\tau {\rm Tr}
\left(F^{\mu \nu} -{i\over 2} \epsilon^{\mu \nu \rho \sigma}F_{\rho \sigma}\right)
({\bar \lambda}_i \tilde{\sigma}_{\nu}\epsilon^i) \cr
&+
{T_3 \over 2}\int d^4 x \sqrt{-g}\, \partial_{\mu}\bar \tau {\rm Tr}
\left(F^{\mu \nu} +{i\over 2} \epsilon^{\mu \nu \rho \sigma}F_{\rho \sigma}\right)
({\bar \lambda}_i \tilde{\sigma}_{\nu}\epsilon^i)+{\rm c.c}.}}
Using the fact that $\tau$ is holomorphic and  that $\epsilon^i$ 
satisfies equations \Aeight, it is easy to see that the first term in the above expression 
vanishes and only the second term containing $\partial_{\mu}\bar \tau$ survives. 
Now we vary the fermionic terms in the action in $S_F$ under:
\eqn\Bfour{
\delta\lambda_{\alpha}^{{\ }i}=-{1\over 2} F_{\mu \nu}
(\sigma^{\mu} \tilde{\sigma}^{\nu})_{\alpha}^{{\ }\beta}\epsilon_{\beta}^{{\ }i}.}
By using the background Killing spinor equation~\zerotwotwelve, we find that 
the terms in $\delta S_F$ with $\partial_{\mu}\tau$ cancel and 
terms with $\partial_{\mu}\bar \tau$ produce exactly the same expression 
as in~\Bthree\ but with the opposite sign. This provides the cancellation 
of terms involving the vectors fields and the fermions. 

Now we consider the variation in the action $S_{Sc}$  involving the scalars. It is straightforward to obtain that:
\eqn\Bfive{\eqalign{
\delta S_{Sc}= &2 T_3 \int d^4 x \sqrt{-g}\, (\partial^{\mu}e^{-\Phi}){\rm Tr}
\left({\cal D}_{\mu}\varphi_{i j}\right) (\lambda^i \epsilon^j) \cr 
&-T_3 \int d^4 x \sqrt{-g}\,e^{-\Phi}\,({\cal R}+\partial^{\mu}\partial_{\mu}\Phi)\, {\rm Tr}
\left(\varphi_{i j}  (\lambda^i \epsilon^j)\right)+ {\rm c. c}.}}
These terms  cancel against the variation of $S_F$ under:
\eqn\Bsix{
\delta {\bar \lambda}_{\dot\alpha i}= 2 i \epsilon^{\alpha j}
\sigma^{\mu}_{{\ }\alpha \dot\alpha}{\cal D}_{\mu}\varphi_{i j}
-{i\over 2}
\epsilon^{\alpha j}
\sigma^{\mu}_{{\ }\alpha \dot\alpha}(\partial_{\mu}\Phi)\varphi_{i j}.}
Let us make some remarks on how the terms containing derivatives of $C$ 
cancel when we vary $S_F$ (such terms are not present in the variation
of $S_{Sc}$ in~\Bfive). Consider the variation of the second term in 
$S_F$ under~\Bsix. We get:
\eqn\Bseven{\eqalign{
&-{i \over 2} \int d^4 x \sqrt{-g}\, \partial_{\mu}C\, {\rm Tr}\left(
{\cal D}_{\nu} \varphi_{i j} 
(\lambda^{i}\sigma^{\mu}\tilde{\sigma}^{\nu}\epsilon^{j})\right) \cr
&+{i \over 8} \int d^4 x \sqrt{-g}\, \partial_{\mu}C\, \partial_{\nu} \Phi\,
{\rm Tr}\left(\varphi_{i j} 
(\lambda^{i}\sigma^{\mu}\tilde{\sigma}^{\nu}\epsilon^{j})\right)+{\rm c. c}.}}
In both terms we   anticommute $\sigma^{\mu}$ and $\tilde{\sigma}^{\nu}$
using~\Athree. Then each term in~\Bseven\ will split into two terms. The first two terms
yield:
\eqn\Beight{
i \int d^4 x \sqrt{-g}\, \partial_{\mu}C\, {\rm Tr}\left(
{\cal D}^{\mu} \varphi_{i j} 
(\lambda^{i}\epsilon^{j})\right)
-{i \over 4} \int d^4 x \sqrt{-g}\, \partial_{\mu}C\, \partial^{\mu} \Phi
{\rm Tr}\left( \varphi_{i j} (\lambda^{i} \epsilon^{j})\right)+{\rm c. c}.}
They cancel against the variation of the fermion kinetic term 
when we rewrite $D \epsilon$ in terms of the derivative of the axion 
by using~\zerotwotwelve. The remaining two terms are:
\eqn\Bnine{\eqalign{
&{i \over 2} \int d^4 x \sqrt{-g}\, \partial_{\mu}C\, {\rm Tr}\left(
{\cal D}_{\nu} \varphi_{i j} 
(\lambda^{i}\sigma^{\nu}\tilde{\sigma}^{\mu}\epsilon^{j})\right) \cr
-&{i \over 8} \int d^4 x \sqrt{-g}\, \partial_{\mu}C\, \partial_{\nu} \Phi
{\rm Tr}\left(\varphi_{i j} 
(\lambda^{i}\sigma^{\nu}\tilde{\sigma}^{\mu}\epsilon^{j})\right)+{\rm c. c}.}}
We now use  the condition that $\tau$ is a holomorphic function together with the projection satisfied by the Killing spinor  
$\tilde{\sigma}_{\bar z}\epsilon^i=0$. This can   be summarized by:
\eqn\Bten{
\partial_{\mu}\tau\, \tilde{\sigma}^{\mu}\epsilon^i=0.}
Using this equation, we can get rid of the terms with derivatives of $C$
in~\Bnine\
and write them using derivatives of $e^{-\Phi}$. The cancellation 
of such terms arising in $\delta S_F$ and $\delta S_{Sc}$ is already straightforward. 

In the last step, we   vary $S_F$ under the remaining term in the variation of $\lambda$:
\eqn\Beleven{
\delta\lambda_{\alpha}^{{\ }i}=-2\, [\varphi_{jk}, \varphi{ki}]\epsilon_{\alpha}^{{\ }j}.}
The terms containing $\partial C$   cancel (after we use the Killing spinor equation $D\epsilon \sim \partial C$
as in~\zerotwotwelve) and we obtain:
\eqn\Btwelve{
-i \int d^4 x \sqrt{-g}\, (\partial_{\mu} e^{-\Phi}){\rm Tr}\left(
[\varphi_{j k}, \varphi^{k i}]({\bar \lambda}_i \tilde{\sigma}^{\mu} \epsilon^j)\right)+{\rm c. c}.}
This term  cancels against the variation of $S_{nab}$. In varying 
$S_{nab}$ we only have to consider: 
\eqn\Bthirteen{
\delta {\bar \lambda}_{\dot\alpha i}=-{i \over 2}
\epsilon^{\alpha j}\sigma^{\mu}_{{\ }\alpha \dot\alpha}(\partial_{\mu}\Phi)\varphi_{k j}.}
Anything else gives terms which cancel just like in flat background. 
It is straightforward to see that the variation of $S_{nab}$ 
under~\Bthirteen\ indeed cancels~\Btwelve. 
This finishes our proof of the supersymmetry of the action.


\appendix{C}{Integrating Out the Defect Fields}

 
In this Appendix, we perform the explicit integration over the defect fields.
We split the $U(N)$ gauge field into an $SU(N)$ gauge field which we denote by $A$ and   
a $U(1)$ gauge field which we denote by $a$. 
Similarly the $U(M)$ gauge field is decomposed into an $SU(M)$ gauge field ${\tilde A}$ and 
a $U(1)$ gauge field ${\tilde a}$.
Therefore, we want to perform the following path 
integral
\eqn\path{
Z=\int [D\chi][D {\bar \chi}][D\tilde{A}][D\tilde{a}] 
\exp\left[\left( (S_{defect}+S_{CS}(\tilde{A})+S_{CS}(\tilde{a})+S_{CS}(a,\tilde{a})\right)\right],}
where:
\eqn\fermions{
S_{defect}=\int dx^+dx^- {\bar \chi}\left(\partial_++A_++\tilde{A}_++a_+ -\tilde{a}_+\right)\chi.}
Here we took into account that $\chi$ carries the opposite $U(1)$ charges under
$U(N)$ and $U(M)$ action.
The non-Abelian  Chern-Simons term $S_{CS}(\tilde{A})$ is given by
\eqn\chern{
S_{CS}(\tilde{A})=-{(2 \pi \alpha^{\prime})^2 \tau_7 \over 2}\int G_5\wedge\hbox{Tr}
\left(\tilde{A}\wedge d\tilde{A}+{2\over 3}\tilde{A}\wedge\tilde{A}\wedge\tilde{A}\right).}
Similarly:
\eqn\cherpnpointone{
S_{CS}(\tilde{a})=
-{(2 \pi \alpha^{\prime})^2 \tau_7 \over 2}\int G_5\wedge
\tilde{a}\wedge d\tilde{a}.}
Finally, the mixed Chern-Simons terms are given by 
\eqn\mixed{
S_{CS}(a,\tilde{a})=-{(2 \pi \alpha^{\prime})^2 \tau_3 \over 2}N
\int G_1\wedge a\wedge \tilde{f} +
{(2 \pi \alpha^{\prime})^2\over 2}M \int G_5\wedge \tilde{a}\wedge  f,}
where $f=da$ and $\tilde{f}=d\tilde{a}$. 

Integrating the fermions in \path\ yields
\eqn\pathsin{
\int [D\chi][D {\bar \chi}] 
\exp\left(i S_{defect}\right)=
\exp{\left[i\left( {M}\Gamma_{WZW}(A)+{N}\Gamma_{WZW}(\tilde{A})+NM\Gamma_{WZW}(a,\tilde{a})\right)\right]}.}
We must now integrate the D7-brane gauge fields $\tilde{A}$ and $\tilde{a}$ in \path. 
The gauge field ${\tilde A}$ is completely decoupled from 
the ${\cal N}=4$ SYM gauge fields $A$ and $a$. Therefore the integral 
over ${\tilde A}$, which appears in the action through the 
terms $N\Gamma_{WZW}(\tilde{A})+S_{CS}(\tilde{A})$ just gives a constant. 

Now  we  have to perform the integral over $\tilde{a}$. In order to simplify the formulas, 
we consider the case of $M$ coincident $D7$-branes with the local $U(1)$ symmetry. 
In this case the RR one-form flux is given by:
\eqn\flux{
G_1={g_s M\over 2\pi}d\theta.}
A similar analysis can be easily generalized for the global solutions, 
as all we require is that $G_1$ satisfies the Bianchi identities. 
The path integral we have to study is 
\eqn\pathtilde{
\int [D\tilde{a}] \exp\left(i \Gamma(a,\tilde{a})\right),}
where
\eqn\actionn{
\Gamma(a,\tilde{a})=N M\Gamma_{WZW}(a-\tilde{a})+ 
S_{CS}(\tilde{a})+ S_{CS}(a,\tilde{a}).}
The explicit expressions are given by
\eqn\WZWsum{
\Gamma_{WZW}(a-\tilde{a})=-{1\over 8\pi}\int dx^+dx^-
\left[ \partial_+\left(u-\tilde{u}\right) \partial_-\left(u-\tilde{u}\right)-
\partial_+\left(u -\tilde{u}\right) \partial_-\left(v -\tilde{v}\right)\right],}
where:
\eqn\gaugeff{
a_+=\partial_+ u, \qquad 
a_-=\partial_+ v, \qquad
\tilde{a}_+=\partial_+ \tilde{u}, \qquad 
\tilde{a}_-=\partial_+ \tilde{v}.}
The Chern-Simons action $S_{CS}(\tilde{a})$ can be simplified to
\eqn\simplychern{
S_{CS}(\tilde{a})=-{1\over 8\pi} N M \int dx^+dx^-d\rho 
\left(\tilde{a}_+\tilde{f}_{-\rho}+
\tilde{a}_-\tilde{f}_{\rho +}+\tilde{a}_\rho\tilde{f}_{+-}\right),}
where $\rho$ is the radial direction away from the $N$ D3-branes and 
we have restricted the RR flux to $s$-waves on the $S^5$.
Likewise
\eqn\mixxed{
S_{CS}(a,\tilde{a})= {1\over 8\pi}N M 
\int dx^+dx^-\left( \tilde{f}_{+-}(0)\int dr\, a_r+ f_{+-}(0)\int d\rho\, \tilde{a}_\rho\right),}
where ${f}_{+-}(0)$ and $\tilde{f}_{+-}(0)$ are the boundary values of 
${f}_{+-}$ and $\tilde{f}_{+-}$ respectively and $r$ is the radial 
coordinate away from the $D7$-branes. 
Note that the path integral is Gaussian and it is enough to evaluate 
the action on the equations of motion. Since we have both bulk and boundary 
contributions to the action we need to solve the equations of 
motion separately on the bulk and on the boundary.

The the bulk equations of motion yield:
\eqn\solveeom{
\tilde{f}_{-\rho}=0, \qquad \tilde{f}_{+\rho}=0, \qquad 2\tilde{f}_{+-}=\tilde{f}_{+-}(0).}
Furthermore, the boundary equations of motion give:
\eqn\boundd{
\int dr\, a_r=-u,\qquad 2\tilde{f}_{+-}(0)= {f}_{+-}(0).}
Evaluating the action on this solution gives:
\eqn\sonshell{
\Gamma(a,\tilde{a})|_{solution}=\Gamma_{WZW}(a).}
Therefore, the final result of performing the path integral \path\ is: 
\eqn\finnn{
Z=\exp\left[i (M\Gamma_{WZW}(A) + M N\Gamma_{WZW}(a))\right].}
We can now combine the $SU(N)$ connection $A$ with the $U(1)$ connection $a$ 
into a $U(N)$ gauge field, which with some abuse of notation we will also denote by $A$.

Therefore, integrating out the localized fields together with the non-dynamical 
gauge fields on the $D7$-branes has the effect of inserting the following 
surface operator into the ${\cal N}=4$ SYM path integral in the $D7$-brane background:
\eqn\insertop{
Z=\exp\left(i M\Gamma_{WZW}(A)\right).}
%


\appendix{D}{A Probe $D7$-Brane in $AdS_5 \times S^5$}


In this Appendix we study the sypersymmetries preserved by the $D7$-brane in $AdS_5 \times S^5$ which represents a surface operator in the probe approximation.

We consider the following parametrization for  $AdS_5\times S_5 $    (we fix the radius $L=1$)
\eqn\ads{ ds^2_{AdS\times S}=\rho^2\eta_{\mu\nu}dx^\mu
dx^\nu+{d\rho^2\over \rho^2} +d\theta^2+\sin^2\theta\;
d\Omega_4^2,}  
where the  metric on $S^4$ is given by: 
\eqn\mq{
d\Omega_4^2=d\varphi_1^2+\sin\varphi_1^2d\varphi_2^2+\sin\varphi_1^2\sin\varphi_2^2d\varphi_3^2+\sin\varphi_1^2\sin\varphi_2^2\sin\varphi_3^2d\varphi_4^2.}
It is useful to introduce tangent space gamma matrices, i.e.
$\gamma_{\underline{m}}=e^m_{\underline{m}}\Gamma_m$
$(m,\underline{m}=0,\ldots,9)$ where $e^m_{\underline{m}}$ is the
inverse vielbein and $\Gamma_m$ are the target space matrices:
 \eqn\flatm{\eqalign{\gamma_\mu={1\over
\rho}\Gamma_\mu\qquad(\mu=0,1,2,3),\qquad\gamma_4=\rho\Gamma_\rho,\qquad\gamma_5=\Gamma_\theta,\cr\gamma_{a+5}={1\over
\sin\theta}\left(\prod_{j=1}^{a-1}{1\over
\sin\varphi_j}\right)\Gamma_{\varphi_{a}}\qquad(a=1,2,3,4)}}
The Killing spinor of $AdS_5\times S^5$ in the coordinates \ads\ is given by~\SkenderisVF\
%
\eqn\ksp{\epsilon=\left[-\rho^{-{1\over
2}}\gamma_4h(\theta,\varphi_a)+\rho^{{1\over
2}}h(\theta,\varphi_a)(\eta_{\mu\nu}x^{\mu}\gamma^{\nu})\right]\eta_2+\rho^{{1\over
2}}h(\theta,\varphi_a)\eta_1}
 where:
\eqn\h{h(\theta,\varphi_a)=e^{{1\over 2}\theta\gamma_{45}}e^{{1\over 2}\varphi_1\gamma_{56}}e^{{1\over 2}\varphi_2\gamma_{67}}e^{{1\over 2}\varphi_3\gamma_{78}}e^{{1\over 2}\varphi_4\gamma_{89}}.}
$\eta_1$ and $\eta_2$ are constant ten dimensional  complex spinors satisfying
%
\eqn\gael{\gamma_{11}\eta_1=-\eta_1\qquad\gamma_{11}\eta_2=\eta_2}
with $\gamma_{11}=\gamma_0\gamma_1\ldots\gamma_9$.  They also satisfy  
\eqn\gas{ \tilde{\gamma}\,\eta_1=\eta_1\qquad
\tilde{\gamma}\,\eta_2=-\eta_2,}
where $\tilde{\gamma}=i\gamma^{0123}$ is the four dimensional chirality matrix.  Thus, each spinor $\eta_{1,2}$ has  16 independent real
components.  

The supersymmetries preserved by the embedding of
a $D$-brane probe, are those that satisfy
\eqn\unsusy{\Gamma_{\kappa}\epsilon=\epsilon,} 
where
$\Gamma_{\kappa}$ is the  $\kappa$-symmetry  transformation matrix 
of the probe worldvolume theory  and $\epsilon$ is the Killing spinor of the
$AdS_5\times S_5$ background \ksp. Both $\Gamma_{\kappa}$ and $\epsilon$
have to be evaluated at the location of the probe. 

Let's now consider  a $D7$-brane  embedding with an $AdS_3\times S^5$ worldvolume geometry,  with embedding:
 \eqn\embds{\eqalign{
&\sigma^0=x^0\qquad \sigma^1=x^1       \qquad \sigma^2=\rho\qquad \sigma^3=\theta  \qquad     \sigma^{3+a}=\varphi^a\qquad (a=1,2,3,4)\cr
& x^2=0\qquad x^3=0}}   
and with the  worldvolume gauge field  set to zero.  The matrix $\Gamma_{\kappa}$ for a $D7$-brane  
in a background with zero $B$-field and dilaton  is given by 
\eqn\fo{d^8\sigma\,\Gamma_{D7}={1\over \sqrt{-\det(g_{ij})}}\,\Gamma_{(8)}I}
where $\Gamma_{(8)}={1\over 8!}\Gamma_{i_1\ldots i_8}d\sigma^{i_1}\wedge\ldots\wedge  d \sigma^{i_8}$   and   $I$ acts on a
spinor $\psi$ by  $I\psi=-i\psi$.  Considering the embedding  in \embds,  the matrix in \fo\ reduces to:
\eqn\kds{\Gamma_{D7}=\gamma_{01456789}I.}

The equation \unsusy\  has to be satisfied at  every point on the worldvolume. Thus, the term proportional to $\rho^{{1\over 2}}$ gives:
\eqn\upm{\Gamma_{D7}h(\theta,\varphi_a)\eta_1=h(\theta,\varphi_a)\eta_1.}
The terms proportional to $\rho^{-{1\over 2}}$, $\rho^{{1\over 2}}x_0$ and $\rho^{{1\over 2}}x_1$  give:
\eqn\upmb{{\Gamma}_{D7}h(\theta,\varphi_a)\eta_2=-h(\theta,\varphi_a)\eta_2.}

Using 
\eqn\grel{h^{-1}\gamma_{014}h=n^I\gamma_{01I}\qquad h^{-1}\gamma_{56789}h=n^I\gamma_{I456789}\qquad I=4,5,6,7,8,9}   where
\eqn\vi{n^I(\theta,\varphi_1,\varphi_2,\varphi_3,\varphi_4)=\left(\eqalign{&\cos\theta\cr&\sin\theta\cos\varphi_1\cr&\sin\theta\sin\varphi_1\cos\varphi_2\cr&\sin\theta\sin\varphi_1\sin\varphi_2\cos\varphi_3\cr&\sin\theta\sin\varphi_1\sin\varphi_2\sin\varphi_3\cos\varphi_4\cr&\sin\theta\sin\varphi_1\sin\varphi_2\sin\varphi_3\sin\varphi_4}\right) } 
is a unit vector in ${\Bbb R}^6$ (that is $n^In^I=1$) we get:
\eqn\hgkh{\eqalign{h^{-1}\Gamma_{D7}h&=n^In^J\gamma_{01I}\gamma_{J456789}I\cr & =-i\gamma_{01456789}\cr &=\gamma^{01}\,\tilde{\gamma}\,\gamma_{11}.}}
Thus, the equations \upm\  \upmb\ reduce to 
\eqn\etacond{\eqalign{\gamma^{01}\eta_1&=-\eta_1\cr  \gamma^{01}\eta_2&=\eta_2}}

Since $\eta_1$  and $\eta_2$  satisfy \gael\ and \gas,  they can be written  in terms of ten dimensional Majorana-Weyl spinors $\epsilon$ and $\varepsilon$ of negative and positive chirality respectively:
\eqn\etr{\eqalign{
\eta_1&=\epsilon+i \gamma^{0123}\epsilon\cr\eta_2&=\varepsilon-i
\gamma^{0123}\varepsilon.}}
By evaluating the Killing spinor \ksp\  near the boundary, $\epsilon$  can be identified with the generator of Poincare supersymmetry while $\varepsilon$ can be identified with the generator of conformal supersymmetry of ${\cal N}=4$ SYM.
Thus the  equations \etacond\ become:
\eqn\etacondf{\eqalign{\gamma^{01}\epsilon&=-\epsilon,\cr  \gamma^{01}\varepsilon&=\varepsilon.}}
These conditions are equivalent to~\poinsusy\ and~\confsusy, which describe the unbroken Poincare and conformal supersymmetries respectively in the field theory.
Therefore, for coincident  $D7$ branes we have shown that they preserve the same half of the Poincare and conformal supersymmetries as the field theory does in the probe approximation.

\subsec{$D7$ probe without conformal supersymmetries}
The $D7$-brane embedding we have just discussed solution  can be generalized to the case when $x^2=\bar{x}^2$ and  $x^3=\bar{x}^3$ where  $\bar{x}^2$ and  $\bar{x}^3$ are arbitrary constants.  The bosonic symmetry of this embedding is $ISO(1,1)\times SO(6)$.
We note that the conformal  and  $U(1)$ symmetries are broken in the case of separated $D7$-branes just like in the field theory.

In this case, the matrix \fo\   is still given by $\kds$.  The supersymmetry conditions are 
\eqn\upmg{\Gamma_{D7}h(\theta,\varphi_a)\eta_1=h(\theta,\varphi_a)\eta_1}
\eqn\upmbg{{\Gamma}_{D7}h(\theta,\varphi_a)\eta_2=-h(\theta,\varphi_a)\eta_2}
\eqn\upme{{\Gamma}_{D7}h(\theta,\varphi_a)\eta_2=h(\theta,\varphi_a)\eta_2.}
The equations \upmbg\ and \upme\ imply that the conformal supersymmetries  are completely broken.  
The equation \upmg\  implies that the preserved Poincare supersymmetries satisfy:     
\eqn\poin{\eqalign{\gamma^{01}\epsilon&=-\epsilon.}}
When $\bar{x}^2=\bar{x}^3=0$  the equation~\upme\ 
does not have to be satisfied and half of the conformal supersymmetries are preserved. 
We thus recover the symmetries preserved by the field theory in the probe approximation.

\lref\GomisIM{
  J.~Gomis and F.~Passerini,
  ``Wilson loops as D3-branes,''
  JHEP {\bf 0701}, 097 (2007)
  [arXiv:hep-th/0612022].
}
\lref\GomisVI{
  J.~Gomis and A.~Kapustin,
  ``Two-dimensional unoriented strings and matrix models,''
  JHEP {\bf 0406}, 002 (2004)
  [arXiv:hep-th/0310195].
}
\lref\GomisBD{
  J.~Gomis and H.~Ooguri,
  ``Non-relativistic closed string theory,''
  J.\ Math.\ Phys.\  {\bf 42}, 3127 (2001)
  [arXiv:hep-th/0009181].
}
\lref\DiaconescuDT{
  D.~E.~Diaconescu and J.~Gomis,
  ``Fractional branes and boundary states in orbifold theories,''
  JHEP {\bf 0010}, 001 (2000)
  [arXiv:hep-th/9906242].
}
\lref\DiaconescuBR{
  D.~E.~Diaconescu, M.~R.~Douglas and J.~Gomis,
  ``Fractional branes and wrapped branes,''
  JHEP {\bf 9802}, 013 (1998)
  [arXiv:hep-th/9712230].
}


\appendix{E}{The Killing Spinor}


The goal of this Appendix is to construct the Killing spinor of the geometry 
dual to the surface operator. The geometry can be written as follows
\eqn\Eone{\eqalign{
& ds^2 =-H_7^{-1/2} H_3^{-1/2}d x^{+} d x^{-} + H_7^{-1/2}H_3^{1/2}d \rho^2
+H_7^{-1/2} d \Omega_5 + H_7^{1/2} H_3^{-1/2} d z d{\bar z}, \cr 
& F_{0123 \rho}=H_7 \partial_{\rho} H_3^{-1},}}
where
\eqn\Etwo{
H_3={L^4 \over \rho^4}}
and $H_7$ is the harmonic function of the $D7$-brane solution. 
To find the Killing spinor we substitute the above solution into the 
gravitino and dilatino variations, which in the presence of one-and five-form 
fluxes take the form: 
\eqn\Ethree{\eqalign{
&\delta\Psi_M=\partial_M \epsilon +{1\over 4}\omega_{M}^{AB}\Gamma_{AB} \epsilon
-{i\over 8}e^{\Phi}\partial_N C \Gamma^N\Gamma_M\epsilon -
{i \over 8 \cdot 5!}e^{\Phi}F_{M_1 \ldots M_5}\Gamma^{M_1 \ldots M_5}\Gamma_{M}\epsilon
=0,\cr 
&\delta \psi =(\Gamma^M\partial_M\Phi)\epsilon 
+i e^{\Phi}\partial_M C \Gamma^M \epsilon=0.}}
The dilatino variation is independent of the five-form flux and gives
\eqn\Efour{
\tau=\tau(z), \qquad \gamma_{\bar z}=0,}
as in the case of $D7$-brane solutions. 
When we substitute~\Eone\ into the gravitino variation, there will be 
terms proportional to $\partial H_7$ and terms proportional $\partial H_3$
which will essentially separate. The term with $\partial H_7$ cancel 
if $\epsilon \sim H_7^{-1/8}$ and~\Efour\ is satisfied 
exactly like 
in the case of $D7$-brane solutions. Let us concentrate on the terms 
proportional to $\partial H_3$. Let us first consider the variation
$\delta \Psi_{z}$. We obtain:
\eqn\Efive{
{H_7^{1/2} \over 8 H_3^{3/2}}\partial_{\rho}H_3 \gamma_{4}\gamma_{z}
(\epsilon +i \gamma_0 \gamma_1 \gamma_2 \gamma_3 \epsilon)=0.}
Note that $\partial_z \epsilon$ cancels against the terms proportional to 
$\partial_z H_7$ and, hence, eq.~\Efive\ is not a differential equation on $\epsilon$. 
To satisfy~\Efive\ we have to require:
\eqn\Esix{
i \gamma^{0 1 2 3}\epsilon =\epsilon.}
Eqs.~\Efour\ and~\Esix\ are equivalent to~\susy\ and \susyz\
and, hence, $\epsilon$ has eight independent components
corresponding to eight preserved supercharges. This is in agreement 
with our field theory discussions. Now we consider the equation
$\delta \Psi_{\pm}=0$. Due to the restriction~\Esix, it follows that 
\eqn\Eseven{
\delta \Psi_{\pm}=\partial_{\pm}\epsilon=0.}
This means that $\epsilon$ is independent of $x^{\pm}$. 
Similarly, from the equation $\delta \Psi_{\rho}=0$ we 
obtain
\eqn\Eeight{
\partial_{\rho}\epsilon-{1 \over 2 \rho}\epsilon =0,}
which implies $\epsilon \sim \rho^{1/2}$. The last equations to consider
is $\delta \Psi_{a}=0$, where $\Psi_{a}$ are the components of the gravitino
along $S^5$. These equations are
\eqn\Enine{
D_a \epsilon-{1 \over 2} \gamma_{4}\Gamma_a \epsilon=0.}
These are the standard equations for the Killing spinor on $S^5$~\Pope, \Claus.
The solution is given in terms of the operator $h(\theta,\varphi_a)$ defined in~\h.
Combining the above conclusions we find that the Killing spinor is given by 
\eqn\Eten{
\epsilon= h(\theta,\varphi_a)H_7^{-1/2}\rho^{1/2}\epsilon_{0},}
where both $\epsilon$ and $\epsilon_0$ satisfy conditions~\susy\ and~\susyz\
(note thate $\gamma_+$ and $\gamma_{\bar z}$ commute with $h(\theta,\varphi_a)$). 


\listrefs

\end

Add to NSERC report \GomisIM\GomisVI\GomisBD\DiaconescuDT\DiaconescuBR